\documentclass[a4paper,10pt]{article}
\pdfoutput=1 

\usepackage{jheppub} 

\usepackage[T1]{fontenc} 
\usepackage{lmodern} 
\usepackage[toc,page]{appendix}

\usepackage{xcolor}
\usepackage{amsmath,amssymb}
\usepackage{mathrsfs,wasysym}
\usepackage{booktabs}
\usepackage{slashed}
\usepackage{cancel}
\usepackage{bm}
\usepackage[normalem]{ulem}

\newcommand{\sss}{\scriptscriptstyle}
\newcommand{\as}{\alpha_s}
\newcommand{\Ord}{\mathcal{O}}
\newcommand{\Lum}{\mathscr{L}}

\newcommand{\muf}{\mu_{\sss\rm F}}
\newcommand{\mur}{\mu_{\sss\rm R}}

\newcommand{\lf}{\log\frac{Q^2}{\mu^2}}

\newcommand{\plus}[1]{\left(#1\right)_+}

\newcommand{\D}{\mathcal{D}}

\newcommand{\Dh}{\hat{\mathcal{D}}}
\newcommand{\Dl}{\mathcal{D}^{\log}}
\newcommand{\Dp}{\hat{\mathcal{D}}^\psi}
\newcommand{\Lb}{\bar L}
\newcommand{\dL}{\delta L}
\newcommand{\dN}{\delta N}
\newcommand{\Li}{\mathrm{Li}}

\let\originalleft\left
\let\originalright\right
\renewcommand{\left}{\mathopen{}\mathclose\bgroup\originalleft}
\renewcommand{\right}{\aftergroup\egroup\originalright}

\def\beq{\begin{equation}}  
\def\eeq{\end{equation}}
\def\({\left(}
\def\){\right)}
\def\[{\left[}
\def\]{\right]}

\let\oldsubsection\subsection
\renewcommand\subsection[2][\subsectiontoc]{%
  \def\subsectiontoc{#2}%
  \oldsubsection[#1]{\boldmath #2}%
}

\let\oldsubsubsection\subsubsection
\renewcommand\subsubsection[2][\subsubsectiontoc]{%
  \def\subsubsectiontoc{#2}%
  \oldsubsubsection[#1]{\boldmath #2}%
}

\allowdisplaybreaks

\title{\boldmath On the approaches to threshold resummation of rapidity distributions for the Drell-Yan process}

\author[a]{Marco Bonvini}
\affiliation[a]{INFN, Sezione di Roma 1,\\ Piazzale Aldo Moro~5, 00185 Roma, Italy}
\author[b]{and Giulia Marinelli}
\affiliation[b]{Universit\`a degli Studi di Milano-Bicocca \& INFN Sezione di Milano-Bicocca, \\ Piazza della Scienza 3, 20126 Milano, Italy}
\preprint{}

\emailAdd{marco.bonvini@roma1.infn.it}
\emailAdd{g.marinelli10@campus.unimib.it}

\abstract{%
  We consider threshold resummation of rapidity distributions, for which various approaches exist in the literature.
  Recently, a work by Lustermans, Michel, Tackmann suggested that older approaches by Becher, Neubert, Xu (BNX)
  and Bonvini, Forte, Ridolfi (BFR) were wrong because they miss some leading power contributions at threshold.
  In this work, we prove and demonstrate that the BNX and BFR approaches
  are correct and able to resum threshold logarithms to leading power accuracy.
  We then show that the BNX and BFR approaches can provide rather good alternatives to more
  modern approaches to threshold resummation of rapidity distributions, provided the threshold logarithms
  are resummed according to the $\psi$-soft definition introduced in the context of Higgs production.
}

\begin{document}

\maketitle

\section{Introduction}
The study of hadron-hadron collisions plays a fundamental role in the understanding
of the Standard Model of particle physics and in the searches for new physics signals beyond it.
Precision is the keyword to fully exploit the potential of hadron-hadron collider machines such as the
Large Hadron Collider (LHC).
Precision must be achieved both at the experimental level and in theoretical predictions.
The latter are the subject of this paper.

Quantum Chromodynamics (QCD) is the main player in the theoretical description of hadron-hadron collisions.
Physical observables at high energy are computed through perturbation theory.
However, in some cases, fixed-order perturbative predictions are not accurate enough.
Indeed, perturbative computations may depend on some logarithms of kinematic origin,
and in some kinematic regions these logarithms can be large and enhanced at every order,
thus spoiling the perturbative result.
In such cases, the all-order resummation of these logarithmic contributions is needed to
stabilise the perturbative result.
It is worth noting that, in some intermediate regions, the same logarithms may be small enough to behave perturbatively
but still large enough to be the dominant contribution to the cross section.
In such cases, the resummation of these contributions is not strictly necessary, but it helps
improving the accuracy of the theoretical prediction.

In this work we consider the threshold region,
usually defined by the limit in which the invariant mass of the tagged final state
is close to the centre-of-mass energy of the initial state.
In this limit, most of the available energy flows into the considered final state,
so that any extra radiation has to be soft.
It is this soft radiation that produces the threshold logarithms whose resummation is the subject of this paper. 

The resummation in the threshold limit is widely discussed in the literature, for a large variety of processes.
The case of the Drell-Yan (DY) pair production is particularly interesting since it is one of the cleanest processes
that can be studied not only to test the Standard Model to a high accuracy,
but also to probe physics beyond the Standard Model.
It has been studied very carefully in perturbative QCD both for inclusive and differential distributions,
such as the rapidity distributions which are considered in this work.
At present, fixed-order perturbative predictions in QCD for the cross section and rapidity distribution of this process
are available up to next-to-next-to-next-leading order (N$^3$LO)~\cite{Duhr:2020seh,Duhr:2020sdp,Chen:2021vtu,Duhr:2021vwj}.

Currently, threshold resummation is well established for inclusive
cross sections~\cite{Catani:1989ne,Sterman:1986aj,Catani:1996yz,Bonciani:1998vc,Forte:2002ni,Becher:2006nr},
while for rapidity distributions there are various approaches considered in the literature
\cite{Becher:2007ty,Bonvini:2010tp,Banerjee:2018vvb,Lustermans:2019cau}
that differ substantially in the way they are constructed.
In particular, the most recent one~\cite{Lustermans:2019cau},
which extends the resummation to a wider kinematic region,
criticises the validity of the first two approaches~\cite{Becher:2007ty,Bonvini:2010tp},
stating that they miss leading contributions.
It is our purpose to analyse this criticism, and show that while the approaches of Refs.~\cite{Becher:2007ty,Bonvini:2010tp}
are certainly less powerful than the one of Ref.~\cite{Lustermans:2019cau},
they are perfectly consistent within their region of validity
and they do not miss any leading contribution at threshold.

It is worth noting that the approach of Ref.~\cite{Banerjee:2018vvb} has been recently extended
to resum also next-to-leading power contributions at threshold~\cite{Ajjath:2021pre},
thereby improving the quality of the result and extending the region of validity of the resummation.
These next-to-leading power contributions are already contained in the approach of Ref.~\cite{Lustermans:2019cau},
which however has not been used to produce all-order results yet.

The structure of this paper is the following.
We give an overview of the state of the art in section~\ref{sec:thr}, also establishing our notation.
We then focus on the approaches of Refs.~\cite{Becher:2007ty,Bonvini:2010tp}
and prove that they are correct within their declared accuracy in section~\ref{sec:correct},
discussing all the criticisms raised in Ref.~\cite{Lustermans:2019cau}.
We assess the quality of these two approaches in section~\ref{sec:good},
and compare them numerically with the approaches of Refs.~\cite{Banerjee:2018vvb,Lustermans:2019cau,Ajjath:2021pre}
in section~\ref{sec:comparison}.
We then present representative resummed results in section~\ref{sec:allorder}.
We conclude in section~\ref{sec:conclusions}.

\section{Threshold resummation of Drell-Yan rapidity distributions}
\label{sec:thr}

We focus on the Drell-Yan process, namely the production of a lepton-antilepton pair
in hadron-hadron collisions.
We denote by $Q^2$ the invariant mass squared of the pair, and by $s$ the collider centre-of-mass energy.
We introduce the variables
\begin{align}\label{eq:tau-z-def}
\tau = \frac{Q^2}{s}, \qquad z = \frac{Q^2}{\hat{s}},
\end{align}
where $\hat{s}=x_1 x_2 s$ is the partonic centre-of-mass energy,
namely the energy of the subsystem identified by the two partons coming from each hadron,
and $x_{1,2}$ are the partons momentum fractions.
According to the QCD factorization theorem, the Drell-Yan cross section
differential in invariant mass squared $Q^2$ and rapidity $Y$ of the pair can be expressed as
\begin{align}\label{eq:XSconv}
\frac{d^2\sigma }{d Q^2 \; dY} = \tau \sigma_0 \sum_{i,j} \int_{\tau}^{1} \frac{dz}{z} \int_{0}^{1} du \; \Lum_{ij} (z,u) \; C_{ij}(z,u,\as)
\end{align}
where $\sigma_0$ is the Born term (given e.g.\ in Ref.~\cite{Anastasiou:2003ds}),
$C_{ij}(z,u,\as)$ are the partonic coefficient functions computable in perturbation theory
and $\Lum_{ij} (z,u)$ are non-perturbative parton luminosities.
We use the variable $u$ in place of the more common parton-level rapidity $y$
(namely the rapidity of the pair defined in the partonic centre-of-mass system)
for later convenience. The two are related by the equations
\begin{align}
\label{eq:u_y}
u= \frac{e^{-2y}-z}{(1-z)(1+e^{-2y})},
\qquad
e^{2y} = \frac{1-(1-z)u}{z+(1-z)u},
\end{align}
which show that the variable $u$ ranges from 0 to 1, as a consequence of the physical constraint $ze^{2|y|}\leq1$.
The parton luminosities are defined as the product of two parton distribution functions
\begin{align}
\label{eq:luminosity}
\Lum_{ij}(z,u)=c_{ij} f_i \(x_1,\muf^2\) f_j \(x_2,\muf^2\)
\end{align}
where the coefficients $c_{ij}$ depend on the vector boson mediating the production of the pair
(see e.g.\ Ref.~\cite{Anastasiou:2003ds} for a general definition) and the momentum fractions
$x_{1,2}$ are given in terms of $z$, $u$, $\tau$ and $Y$ by
\begin{subequations}\label{eq:x12}
\begin{align}
x_1 &= \sqrt{\frac\tau z}e^{Y-y} = \sqrt{\frac\tau z}e^Y \sqrt{\frac{z+(1-z)u}{1-(1-z)u}} , \\
x_2 &= \sqrt{\frac\tau z}e^{y-Y} = \sqrt{\frac\tau z}e^{-Y} \sqrt{\frac{1-(1-z)u}{z+(1-z)u}} .
\end{align}
\end{subequations}
The physical constraint $x_{1,2} \leq 1$ restricts the actual integration range of $u$, giving effectively
\begin{align}\label{eq:ulimits}
\max\[0,\frac{\tau e^{-2Y}-z^2}{(1-z)(\tau e^{-2Y}+z)}\] \leq u \leq \min\[1,\frac{z(1-\tau e^{2Y})}{(1-z)(\tau e^{2Y}+z)}\].
\end{align}
For a lighter notation, we are omitting the explicit dependence on $Y$, $\tau$ and on the factorization scale $\muf$ of the luminosity,
which is always implicitly understood.
Similarly, we are not showing the dependence on the factorization and renormalization scales of the partonic coefficient function,
as their dependence is not central in our discussion.

\subsection{The threshold limit(s)}

We now consider the threshold limit. As already mentioned in the introduction,
this is the limit in which the available energy is just enough to produce the tagged final state.
Namely, Drell-Yan is at threshold when the collider energy $\sqrt s$ is close to the invariant mass $Q$ of the lepton pair,
or in terms of the variables defined in Eq.~\eqref{eq:tau-z-def} when $\tau\to1$ (from below, as $\tau$ has to be smaller than 1 by definition).
This limit, called henceforth the \emph{hadronic threshold limit}, is not very interesting phenomenologically
because the cross section is small (at $\tau=1$ it becomes identically zero).

A more interesting limit is the so-called \emph{partonic threshold limit},
where it is the parton level centre-of-mass energy $\sqrt{\hat s}$ which is close to $Q$,
namely $z\to1$.
In the cross section formula Eq.~\eqref{eq:XSconv}, the variable $z$ is integrated over in the range $\tau\leq z\leq1$.
Consequently, if $\tau$ is close to 1 then also $z$ is forced to be close to 1, namely hadronic threshold implies partonic threshold.
However, the converse is not true: even far from hadronic threshold (which is the phenomenologically interesting region for Drell-Yan)
contributions from the partonic threshold region $z\to1$ are always present, as the integral always extends to $z=1$.
Most importantly, the partonic threshold region often dominates the integral,
due to the shape of the PDFs that act as a weight favouring large values of $z$ even when $\tau$ is small.
This phenomenon is sometimes called \emph{dynamical threshold enhancement}~\cite{Becher:2007ty},
and it has been quantified through a saddle point argument in Mellin space~\cite{Bonvini:2010tp,Bonvini:2012an}
both in the context of Drell-Yan and of Higgs production.

The partonic threshold region is relevant also because the threshold logarithms
mentioned in the introduction appear in the perturbative computations as logarithms in the variable $z$.
More specifically, the partonic coefficient function in the $q\bar q$ channel 
develops contributions of the form (the $+$ suffix denotes the usual plus distribution)
\beq\label{eq:thrlogs}
\as^n \plus{\frac{\log^k (1-z)}{1-z}},
\qquad 0\leq k< 2n,
\eeq
at order $n$ in the strong coupling $\as$.
Because the highest power of the logarithm grows with the order, with two extra power for each extra order,
the coefficient function is said to be affected by a double logarithmic enhancement.
Clearly, in the partonic threshold limit, these enhanced logarithms spoil the reliability of
a fixed-order computation and need to be resummed to all orders.

On top of these singular contributions in the $z$ variables there are other plus distributions
and Dirac delta terms in the variable $u$, that are singular in $u=0$ or $u=1$.
These contributions are not directly related to the threshold region,
but they play a role in the accurate description of the parton-level coefficient function
at the partonic rapidity endpoints. We will come back to this point later in this section.

Because threshold logarithms appear only in the $q\bar q$ channel,\footnote
{Contributions from other channels are suppressed by at least one power of $1-z$ with respect to the threshold logarithms Eq.~\eqref{eq:thrlogs}.}
from now on we consider only its
contribution to the cross section. We write the $q\bar q$ contribution to Eq.~\eqref{eq:XSconv} as
\begin{align}\label{eq:XSconv2}
\frac{d^2\sigma_{q\bar q}}{d Q^2 \; dY} = \tau \sigma_0 \int_{\tau}^{1} \frac{dz}{z} \int_{0}^{1} du \; \Lum_{q\bar q} (z,u) \; C(z,u,\as)
\end{align}
in terms of the ``total'' $q\bar q$ luminosity defined by
\beq
\label{eq:luminosityqq}
\Lum_{q\bar q}(z,u)=\sum_q c_{q\bar q} f_q \(x_1,\muf^2\) f_{\bar q} \(x_2,\muf^2\)
\eeq
and where we have removed the subscript $q\bar q$ from the coefficient function $C(z,u,\as)$ to keep the notation light.

As long as rapidity-integrated distributions are concerned (e.g.\ the invariant mass distribution or the total cross section)
the definition of the threshold logarithm Eq.~\eqref{eq:thrlogs} is unique.
However, when considering rapidity distributions, it is possible to distinguish between
the logarithms originating from each incoming quark.
To do so, it is more convenient to use a different set of variables: in place of $z,u$ (or $z,y$)
one can use $z_a,z_b$ related to the former by
\begin{subequations}\label{eq:zazbdef}
\begin{align}
  z_a&=\sqrt z e^y = \sqrt z \sqrt{\frac{1-(1-z)u}{z+(1-z)u}}, \\
  z_b&=\sqrt z e^{-y} = \sqrt z \sqrt{\frac{z+(1-z)u}{1-(1-z)u}},
\end{align}
\end{subequations}
 which in turn gives $z=z_az_b$. Each of these variables is related to each incoming parton.
In particular, the momentum fractions $x_{1,2}$ defined in Eq.~\eqref{eq:x12} are given by
\begin{subequations}
\begin{align}
  x_1&=\frac{x_a}{z_a}, & x_a&=\sqrt\tau e^Y, \\
  x_2&=\frac{x_b}{z_b}, & x_b&=\sqrt\tau e^{-Y},
\end{align}
\end{subequations}
with $x_ax_b=\tau$.
In terms of these variables Eq.~\eqref{eq:XSconv2} takes the form of a double Mellin convolution,
\begin{align}\label{eq:XSconv3}
  \frac{d^2\sigma_{q\bar q}}{d Q^2 \; dY} = \tau \sigma_0 \int_{x_a}^{1} \frac{dz_a}{z_a} \int_{x_b}^{1} \frac{dz_b}{z_b} \; \tilde C\(z_a,z_b,\as\)
  \sum_q c_{q\bar q} f_q \(\frac{x_a}{z_a},\muf^2\) f_{\bar q} \(\frac{x_b}{z_b},\muf^2\),
\end{align}
where
\beq
\tilde C\(z_a,z_b,\as\) = \frac{dz\, du}{dz_a\,dz_b}C\(z(z_a,z_b),u(z_a,z_b),\as\).
\eeq
The coefficient function in terms of these variables contains double logarithms of the form Eq.~\eqref{eq:thrlogs}
but in the variables $z_a$ and $z_b$ separately:
\beq\label{eq:thrlogs2}
\as^n \plus{\frac{\log^k (1-z_a)}{1-z_a}}
\qquad \text{and} \qquad
\as^n \plus{\frac{\log^k (1-z_b)}{1-z_b}},
\qquad 0\leq k< 2n.
\eeq
These are related to the threshold logarithms in the variable $z$,
but the conversion is not straightforward, as it involves also the $u$ dependence (see Ref.~\cite{Lustermans:2019cau} for more detail).
For completeness, we report in appendix~\ref{sec:NLO} the NLO contribution to the coefficient function,
written with the two choices of sets of variables.

When using these parton-specific variables, there are two regions that generate large logarithms:
$z_a\to1$ and $z_b\to1$.
The threshold region discussed before, $z\to1$, coincides with the overlap of the two regions $z_a\to1$ and $z_b\to1$,
because of the relation $z=z_az_b$.
However, if just one of these two variables is large, say $z_a\to1$, and the other is not large, $z_b\ll1$,
then the variable $z$ is also not large and so the threshold logarithms Eq.~\eqref{eq:thrlogs} are harmless,
but there are large logarithms in the coefficient functions (those from $z_a$) which may spoil perturbativity.
This mechanism can be understood in terms of the $z,u$ variables noting that
there are other large contributions in the coefficient function
coming from the $u$ dependence, and in particular at the edge of the range of definition of $u$.
Indeed, it is immediate to see from Eq.~\eqref{eq:zazbdef} that $z_a\to1$ corresponds to $u\to0$, and $z_b\to1$ corresponds to $u\to1$.
These singular contributions from the $u$ dependence are enhanced at the partonic rapidity endpoints,
as one can see from Eq.~\eqref{eq:u_y}, and are therefore relevant to describe the tails of the rapidity distribution.

These considerations show that there is a region, larger than the partonic threshold region previously defined,
where logarithms of the form Eq.~\eqref{eq:thrlogs2} may become large, possibly spoiling the reliability of the perturbative result.
Adopting a notation introduced in Ref.~\cite{Lustermans:2019cau}, we may call it \emph{generalized partonic threshold region}.
We can see this as the partonic version of the generalized hadronic threshold region of Ref.~\cite{Lustermans:2019cau},
identified by the condition $\tau e^{2|Y|}\to 1$ (or equivalently either $x_a$ or $x_b$ close to 1, depending on the sign of $Y$),
which corresponds to the tails of the rapidity distribution irrespectively of the value of $\tau$.
In other words, it corresponds to the region in which the production of the Drell-Yan pair is at threshold at a given value of the rapidity $Y$.
In this region, either $z_a$ or $z_b$ is forced to be large,
but the other variable can take any accessible value.
Therefore, the generalized hadronic threshold limit implies the generalized partonic threshold limit.
It is worth noting that in the stronger hadronic threshold limit $\tau\to1$, both $z_a$ and $z_b$ are forced to be large,
and therefore this generalized partonic threshold region coincides with the partonic threshold region $z\to1$.

\subsection{Threshold resummation}

Large logarithms in the coefficient function need to be resummed in order to obtain reliable predictions.
In the literature there exist at least two families of approaches:
one aiming at resumming the logarithms in the variable $z$, Eq.~\eqref{eq:thrlogs},
and one aiming at resumming the logarithms in the variables $z_a$ and $z_b$, Eq.~\eqref{eq:thrlogs2}.
As we stressed already,
the second family is more powerful as it resums more terms than what is resummed in the first family,
making the resummed result useful in a wider kinematic range.

To our knowledge, there exist four approaches to resum threshold logarithms in rapidity distribution for the Drell-Yan process.
We present them in chronological order of appearance.
\begin{itemize}
\item The approach of Becher, Neubert, Xu (BNX henceforth)~\cite{Becher:2007ty}, belonging to the first family.
  It is based on the observation that the $u$ dependence in the PDF luminosity Eq.~\eqref{eq:luminosityqq}
  is next-to-leading power at large $z$, Eq.~\eqref{eq:x12}, which allows to write the leading power
  contribution in terms of the rapidity-integrated coefficient function, whose resummation is well known.\footnote
  {In particular, BNX use SCET to compute the resummed coefficient function, but different choices are possible
    without affecting the structure of the resummed formula for the rapidity distributions.}
\item The approach of Bonvini, Forte, Ridolfi (BFR henceforth)~\cite{Bonvini:2010tp}, belonging to the first family.
  It is similar to the BNX approach, but the derivation is based on a different argument in Mellin-Fourier space~\cite{Bolzoni:2006ky},
  and extends the older result of Ref.~\cite{Laenen:1992ey}.
  The resulting resummation formula differs from BNX, but it has been shown to be equivalent up to next-to-next-to-leading power in $1-z$.
\item The approach of Banerjee, Das, Dhani, Ravindran (BDDR henceforth)~\cite{Banerjee:2018vvb} (see also~\cite{Das:2023bfi}), belonging to the second family.
  This is a two-scale extension of the original approaches to rapidity-integrated resummation~\cite{Catani:1989ne,Sterman:1986aj},
  already introduced for $x_F$ distributions in Ref.~\cite{Catani:1989ne}
  and later converted to rapidity distributions in Refs.~\cite{Mukherjee:2006uu,Ravindran:2006bu,Ravindran:2007sv,Westmark:2017uig,Banerjee:2017cfc},
  where a double Mellin transform is taken with respect to the two variables $z_a$ and $z_b$ and logarithms of the product
  of the two Mellin conjugate variables are resummed to all orders.
\item The approach of Lustermans, Michel, Tackmann (LMT henceforth)~\cite{Lustermans:2019cau}, belonging to the second family.
  It uses the framework of soft-collinear effective theory to factorise the cross section in terms of beam functions and a soft function,
  enabling the resummation of threshold logarithms after solving renormalization group equations that they obey.
  It is designed to be valid in the generalized threshold region, thus enlarging its kinematic range of applicability.
\end{itemize}
We stress that the BDDR approach has been recently extended~\cite{Ajjath:2021pre,Ahmed:2020amh,Ajjath:2022kyb,Ajjath:2020sjk}
to resum next-to-leading power contributions to the dominant $q\bar q$ channel,
namely those suppressed by one power of $1-z_a$ or $1-z_b$ with respect to the leading power logarithms Eq.~\eqref{eq:thrlogs2}.
This extension allows to enlarge the region where threshold contributions dominate, and is therefore very useful
to obtain reliable predictions close to threshold and more precise predictions even far from threshold,
even though for achieving higher accuracy the next-to-leading power contributions from other channels,
  i.e.\ the $qg$ channel, should be included as well.

We also stress that the LMT approach already captures the subleading power contributions
predicted in Refs.~\cite{Ajjath:2021pre,Ahmed:2020amh,Ajjath:2022kyb,Ajjath:2020sjk}.
Indeed, it is constructed to resum the leading power terms in one of the two variables, say $z_a$,
with full dependence on the other variable, $z_b$, and vice versa.
Therefore, it contains contributions suppressed with \emph{any} power of $1-z_b$ which multiples the
leading-power terms in $z_a$, and vice versa.\footnote
{These towers of contributions are referred to as leading power in the generalized threshold limit (LP$_{\rm gen}$) in Ref.~\cite{Lustermans:2019cau}.}
Moreover, the LMT approach is able to predict also the off-diagonal $qg$ channel at this accuracy.
We recall that the current LMT work~\cite{Lustermans:2019cau} presents only the analytic structure of this resummation
but it contains no all-order numerical results yet.

The LMT paper, on top of proposing the virtually best approach to resum threshold logarithms in rapidity distributions,
criticises the BNX and BFR approaches, stating explicitly that they are wrong as they miss leading power contributions in $1-z$.
We disagree with this criticism, and we will show in the next section that the BNX and BFR approaches,
which have been used in the literature~\cite{Bonvini:2013jha,Bonvini:2015ira,Pecjak:2018lif,Bargiela:2022dla},
are perfectly consistent within their region of validity.

\section{On the validity of BNX and BFR}
\label{sec:correct}

In this section we focus on the approaches of BNX and BFR,
whose validity has been criticised in Ref.~\cite{Lustermans:2019cau}.
We will present a detailed proof valid for both approaches,
that also reveals the expected quality of the threshold approximation at the core of these approaches.
We discuss possible caveats, and finally comment on the arguments
of Ref.~\cite{Lustermans:2019cau} against the validity of BNX and BFR.

Before moving to this, we recall the explicit expressions of BNX and BFR resummation,
also to establish our notation.
When using the variables $z$ and $u$, in the partonic threshold region $z\to1$ the coefficient function
can be expanded as (we omit the argument $\as$ from now on to emphasise the dependence on the other variables)
\beq
\label{eq:factorizedC_thr}
C(z,u) = C_{\rm thr}(z,u)\, \[1+\Ord(1-z)\],
\eeq
where $C_{\rm thr}(z,u)$ is defined to contain all leading power threshold contributions, namely the plus distributions Eq.~\eqref{eq:thrlogs}
and delta functions of $1-z$.\footnote
{Note that there may be terms that are apparently singular in $z=1$ but multiply some $u$-dependent contribution
that makes them integrable. In this case, the plus distribution is not needed, see e.g.\ the first term in the second line of Eq.~\eqref{eq:C1_NLO}.
We do not consider these terms as leading power, while they are considered such in Ref.~\cite{Lustermans:2019cau}.
This is part of the LMT criticism to the BNX and BFR approaches, and we will discuss it in detail in section~\ref{sec:leadingpower}.}
In the BNX approach the $u$ dependence of the coefficient function at threshold is further approximated as
\beq
\label{eq:BNX}
C_{\rm thr}^{\rm BNX} (z,u)\equiv \frac{\delta(1-u) + \delta(u)}{2} C_{\rm thr}(z),
\eeq
while BFR find
\beq
\label{eq:BFR}
C_{\rm thr}^{\rm BFR} (z,u) \equiv \delta \( u - \frac12\) C_{\rm thr}(z),
\eeq
where $C_{\rm thr}(z)$ is the rapidity-integrated coefficient function at threshold,
reported at fixed order in appendix~\ref{sec:approx}.
Indeed, it is immediate to see that the integral over $u$ of both expressions gives exactly $C_{\rm thr}(z)$.
This coefficient function is then the threshold limit of the well known inclusive coefficient function,
whose resummation has been studied for
decades~\cite{Catani:1989ne,Sterman:1986aj,Catani:1996yz,Bonciani:1998vc,Forte:2002ni,Becher:2006nr}.
We stress that in the original BNX and BFR approaches two different methods for resumming $C_{\rm thr}(z)$ were considered,
but this is immaterial as any other choice is possible.
Therefore, for our purposes, we refer to BNX and BFR as the two formulas Eq.~\eqref{eq:BNX} and \eqref{eq:BFR},
irrespectively of how $C_{\rm thr}(z)$ is resummed to all orders.

\subsection{Proof of BNX and BFR}
\label{sec:proof}

The approximations Eq.~\eqref{eq:BNX} and Eq.~\eqref{eq:BFR} are not particularly significant written in that way.
Indeed, the $u$ dependence of the coefficient function is distributional, and it is clear for instance
that each equation cannot be seen as an approximation of the other, namely there is no sense in which $\delta(u)+\delta(1-u)\simeq2\delta(u-\frac12)$.
The meaning of Eq.~\eqref{eq:BNX} and Eq.~\eqref{eq:BFR} resides in the way they appear in the
cross section formula, Eq.~\eqref{eq:XSconv2}, where also the parton luminosity is present
and plays a crucial role in constructing the BNX/BFR approximations.
We now show this in detail.

The key observation is the fact that the PDFs become independent of $u$ in the partonic threshold limit $z\to1$.
This can be seen by expanding the arguments of the PDFs, Eq.~\eqref{eq:x12}, in powers of $1-z$:
\begin{subequations}\label{eq:x12exp}
\begin{align}
x_1 &= \sqrt{\tau}e^Y \[1+u(1-z) + \Ord((1-z)^2)\],\\
x_2 &= \sqrt{\tau}e^{-Y} \[1+(1-u)(1-z) + \Ord((1-z)^2)\].
\end{align}
\end{subequations}
Consequently, the luminosity Eq.~\eqref{eq:luminosityqq} can be expanded as
\beq\label{eq:lumexp}
\Lum_{q \bar{q}}(z,u) = \Lum_{q \bar{q}}(1,u) - \Lum'_{q \bar{q}}(1,u)(1-z) + \Ord[(1-z)^2],
\eeq
with $\Lum'_{q\bar q}$ the derivative with respect the $z$ variable,
where the first term which is the luminosity computed in $z=1$ is independent of $u$,
\beq
\Lum_{q \bar{q}}(1,u) = \sum_qc_{q \bar{q}} f_q\(\sqrt{\tau}e^Y,\muf^2\) f_{\bar{q}}\(\sqrt{\tau}e^{-Y},\muf^2\).
\eeq
When we will want to emphasise this independence, we will write it as $\Lum_{q \bar{q}}(1,\cdot)$.
Note that the second term of Eq.~\eqref{eq:lumexp} can be considered as truly suppressed by a power of $1-z$
if $\Lum_{q\bar q}'(1,u)$ is of the same size of $\Lum_{q \bar{q}}(1,\cdot)$.
This is not always the case.
We will investigate numerically the size of $\Lum_{q\bar q}'(1,u)$,
which may potentially be a limiting factor of our proof of BNX and BFR,
in section~\ref{sec:limitations}.

The fact that the $u$ dependence of the luminosity is power suppressed in the partonic threshold limit $z\to1$
shows that the $u$ dependence of the coefficient function can be integrated over at leading power.
To formally prove this, we consider the cross section formula Eq.~\eqref{eq:XSconv2} and
we expand the luminosity at threshold,
along the lines of Ref.~\cite{Becher:2007ty}:
\begin{align}
\label{eq:proof}
\frac{1}{\tau\sigma_0} \frac{d^2 \sigma_{q \bar{q}} }{d Q^2 \; dY}
&= \int_\tau^1 \frac{dz}z \int_{0}^{1} du\; \[\Lum_{q \bar{q}}(1,\cdot) + \Ord(1-z) \]\, C(z,u) \nonumber\\
&= \Lum_{q \bar{q}}(1,\cdot) \int_\tau^{1} \frac{dz}z \int_{0}^{1} du\; C(z,u)\, \[1+\Ord(1-z)\] \nonumber\\
&= \Lum_{q \bar{q}}(1,\cdot) \int_\tau^{1} \frac{dz}z \; C(z)\, \[1+\Ord(1-z)\] \nonumber\\
&= \int_\tau^{1} \frac{dz}z \; \Lum_{q \bar{q}}(z,\bar u)\, C(z)\, \[1+\Ord(1-z)\] \nonumber\\
&= \int_\tau^{1} \frac{dz}z \; \Lum_{q \bar{q}}(z,\bar u)\, C_{\rm thr}(z)\, \[1+\Ord(1-z)\].
\end{align}
Here we have first used Eq.~\eqref{eq:lumexp} to expand the luminosity at large $z$,
then we used the $u$-independence of the luminosity in $z=1$ to pull it out of the integral,
we then computed the $u$ integral to obtain the rapidity-integrated coefficient function $C(z)$.\footnote
{Note that in this step we also used the fact that in the $z\to1$ limit the whole region $0\leq u\leq1$ is kinematically allowed,
  as a consequence of the fact that the restriction imposed by the condition $x_{1,2}<1$, expressed by the limits Eq.~\eqref{eq:ulimits},
  are immaterial in the $z\to1$ limit.
  This is also obvious from the fact that the restriction on $u$ would be imposed by $\theta$ functions hidden
  in the luminosity, which in $z=1$ no longer depends on $u$.}
In the penultimate line we restored the $z$ dependence of the luminosity, which is again legitimate up to $\Ord(1-z)$
thanks to Eq.~\eqref{eq:lumexp}, but in doing so we
also need to restore the $u$ dependence in the second argument.
As $u$ has been integrated over, any value $\bar u$ between 0 and 1 is formally acceptable.
Finally, in the last step we have further approximated $C(z)$ with its threshold limit.

We stress that in Eq.~\eqref{eq:proof} we have expanded in powers of $1-z$ selected terms of the \emph{entire} integrand.
This is in contrast with the standard approach of threshold resummation~\cite{Catani:1989ne,Sterman:1986aj},
where only the coefficient function is expanded.
Keeping the parton luminosity unexpanded is certainly more natural
  and ensures for instance that kinematic constraints coming from the limits of PDF arguments
  are preserved by the resummed expression.
  However, from a formal point of view, it is clearly allowed to also include the parton luminosity in the threshold expansion.

The selection of terms that are expanded in Eq.~\eqref{eq:proof} is obviously
arbitrary but legitimate according to the power counting in $1-z$,
and it is made in a way to reproduce BNX and BFR as we shall see.

In the derivation of Eq.~\eqref{eq:proof} we have made an assumption that is not entirely trivial.
In particular, in the second step we have assumed the identity
\beq\label{eq:lumid}
\Lum_{q \bar{q}}(1,\cdot) + \Ord(1-z) = \Lum_{q \bar{q}}(1,\cdot) \,\[1+\Ord(1-z)\].
\eeq
As mentioned above,
this assumption is correct if $\Lum_{q\bar q}'(1,u)$
is of the same size of $\Lum_{q \bar{q}}(1,\cdot)$, which is not always the case.
By comparing numerically the size of the two functions, we will see in section~\ref{sec:limitations}
that this assumption is satisfied in a wide kinematic range.
In particular, at small $\tau$ and not too close to the rapidity endpoints,
which is the phenomenologically relevant region, the assumption Eq.~\eqref{eq:lumid}
is valid and Eq.~\eqref{eq:proof} is correct.

Note that the contributions of $\Ord(1-z)$ in the integrand of Eq.~\eqref{eq:proof} may be large,
as the integral extends down to $z=\tau$ and if $\tau$ is small there is a part of the integration region
in which these corrections are not suppressed.
This fact does not invalidate the derivation of Eq.~\eqref{eq:proof},
but poses questions about the quality of an approximation obtained neglecting those subleading power terms.
We will see in section~\ref{sec:limitations}
that the approximation for the dependence on the luminosity is rather good even for small values of $\tau$.
The impact of approximating $C(z)$ with $C_{\rm thr}(z)$ in the last step
of Eq.~\eqref{eq:proof} depends on the form of threshold logarithms used to construct it,
and it will be discussed in section~\ref{sec:good}.

Having established Eq.~\eqref{eq:proof}, we can obtain the BNX and BFR formulation
by neglecting the subleading power contributions of $\Ord(1-z)$ and for suitable choices of the variable $\bar u$.
In particular, we obtain the BFR formulation by choosing $ \bar{u} = \frac{1}{2}$,
while the BNX one corresponds to the average of the result with $\bar{u} = 0$ and $\bar{u}=1$:
\begin{align}
  \label{eq:BNX_form}
\frac{1}{\tau\sigma_0} \frac{d^2 \sigma^{\rm BNX}_{q \bar{q}} }{d Q^2 \; dY} &= \int_{\tau}^{1} \frac{dz}z \; \frac{ \Lum_{q \bar{q}}(z,0) + \Lum_{q \bar{q}}(z,1)}{2} C_{\rm thr}(z),\\
	\label{eq:BFR_form}
\frac{1}{\tau\sigma_0} \frac{d^2 \sigma^{\rm BFR}_{q \bar{q}} }{d Q^2 \; dY} &= \int_{\tau}^{1} \frac{dz}z \; \Lum_{q \bar{q}}\bigg( z, \frac{1}{2} \bigg) C_{\rm thr}(z). 
\end{align}
It is immediate to verify that these are indeed the results obtained by using Eq.~\eqref{eq:BNX} and Eq.~\eqref{eq:BFR} into Eq.~\eqref{eq:XSconv2}.
In order to use these equations for resummation,
  the function $C_{\rm thr}(z)$ needs to be resummed to all orders in the threshold limit.
  The resummed result should then be matched to fixed-order computations, which shall not be approximated.
  The quality of the resummed and matched result will also depend on whether the resummed threshold logarithms
  are the dominant part of the higher orders or not: we will address this question in section~\ref{sec:good}.

Note that BNX and BFR are just two of infinitely many possible alternative and equivalent formulations of resummation,
which can be obtained by using different values of $\bar u$ and averages thereof.
We notice however that not all combinations make physical sense.
Indeed, the rapidity distribution for the Drell-Yan process in proton-proton collisions
is forward-backward symmetric, but the luminosity $\Lum(z,u)$ is not symmetric for $Y\to1-Y$ unless $u=1/2$.
More precisely, the luminosity Eq.~\eqref{eq:luminosityqq} is symmetric under the exchange of $x_1$ and $x_2$,
which can in turn be realised by changing $Y\to-Y$ and $u\to1-u$ simultaneously, see Eq.~\eqref{eq:x12}.
Therefore, only symmetric sums $\Lum(z,u)+\Lum(z,1-u)$ are symmetric in $Y$.
Since the dependence on the rapidity $Y$ of Eq.~\eqref{eq:proof} only comes from the luminosity,
the general physically acceptable resummed expression is given by
\begin{align}
\label{eq:XSproven}
\frac{1}{\tau\sigma_0} \frac{d^2 \sigma_{q \bar{q}}^{\rm res} }{d Q^2 \; dY}
&= \int_\tau^{1} \frac{dz}z \; \frac{\Lum_{q \bar{q}}(z,\bar u)+\Lum_{q \bar{q}}(z,1-\bar u)}2 C_{\rm thr}(z),
\end{align}
where any value of $\bar u$ in the allowed range $0\leq\bar u\leq1$ is acceptable.
Also any weighted average of results with different $\bar u$ provides a valid formulation of resummation.
Obviously, BNX and BFR are two (maximally different) special cases of Eq.~\eqref{eq:XSproven}.

We recall that the integrands of the BFR and BNX expressions differ effectively by $\Ord[(1-z)^2]$ terms,
as it was noticed in Ref.~\cite{Bonvini:2010tp}.
This fact does not mean that the accuracy of these two formulations is higher, it just tells us that
the two approaches are more similar than expected from the accuracy of the derivation.
We will see this explicitly in the numerical results of the next sections.
In fact, it is easy to prove that any symmetric average of the form Eq.~\eqref{eq:XSproven}
with any value of $\bar u$
is equivalent to BFR and BNX up to $\Ord[(1-z)^2]$.

\subsection{On the validity of the threshold expansion in the integrand}
\label{sec:limitations}

The derivation of the BNX and BFR results in Eq.~\eqref{eq:proof}
involves a number of assumptions and expansions whose validity we now address.

We have already commented that the derivation of Eq.~\eqref{eq:proof} assumes
the equality Eq.~\eqref{eq:lumid}, which in turn is valid if $\Lum_{q\bar q}'(1,u)$
is of the same size of $\Lum_{q \bar{q}}(1,\cdot)$.
This is usually the case, but when the PDFs are computed close to their endpoint $x=1$,
which happens when $\tau e^{2|Y|}$ is close to 1,
this assumption is no longer valid. We now investigate this effect in detail.

Let us focus on the case in which $\tau$ is close to 1
(the case in which $\tau$ is not large but the rapidity is large is similar but slightly more complicated to describe analytically).
In this case, the identity Eq.~\eqref{eq:lumid} is no longer valid.
We can see this analytically, by approximating the PDFs as
\beq\label{eq:PDFtoy}
f(x,\muf^2)\simeq (1-x)^\alpha
\eeq
for some positive value of $\alpha$, which is a good approximation at large $x$.
Considering a single flavour for simplicity (or equivalently assuming that the same value of $\alpha$ holds for all quark PDFs),
the derivative of the luminosity can be written as
\beq
\Lum_{q\bar q}'(1,u) \simeq -\alpha\sqrt\tau \( \frac{e^Yu}{1-\sqrt\tau e^Y} + \frac{e^{-Y}(1-u)}{1-\sqrt\tau e^{-Y}}\)\Lum_{q \bar{q}}(1,\cdot).
\eeq
It is clear that at large $\tau e^{2|Y|}\to1$ one of the two denominators becomes parametrically small,
and thus for some values of $u$ the derivative becomes parametrically larger than the luminosity.
This becomes even clearer at central rapidity $Y=0$,
\beq
\Lum_{q\bar q}'(1,u) \overset{Y=0}\simeq -\alpha \frac{\sqrt\tau+\tau}{1-\tau}\Lum_{q \bar{q}}(1,\cdot),
\eeq
where the denominator clearly enhances the derivative with respect to the luminosity for any value of $u$
in the $\tau\to1$ limit.
In the derivation of Eq.~\eqref{eq:proof}, therefore, when we assume that $\tau$ is close to 1
which in turn implies that $1-z$ is of the same order as $1-\tau$,
the term $-\Lum_{q\bar q}'(1,u)(1-z)$ is not subleading power
with respect to $\Lum_{q \bar{q}}(1,\cdot)$ in Eq.~\eqref{eq:lumexp} and the expansion in Eq.~\eqref{eq:proof} is not accurate.

\begin{figure}[t]
  \centering
  \includegraphics[width=\textwidth]{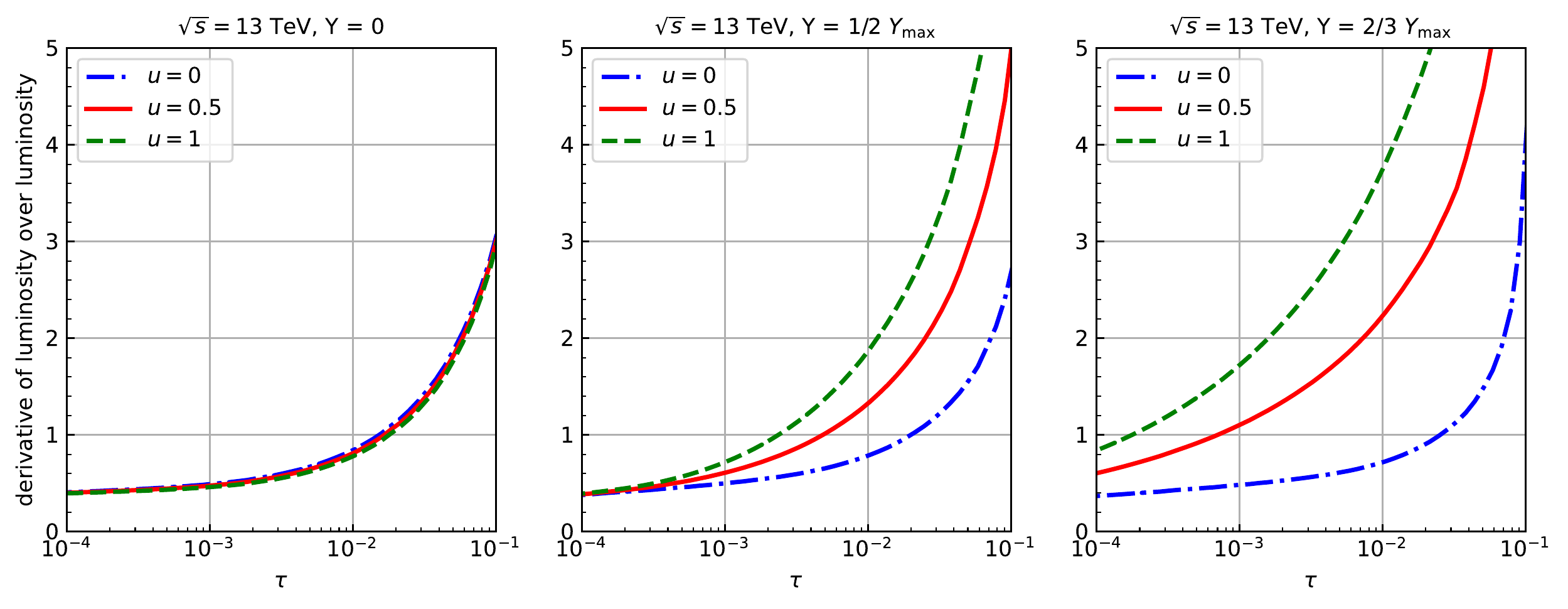}
  \caption{Ratio of the derivative of the general luminosity over the luminosity in $z=1$
    as a function of $\tau$ for $Y = 0$ (left), $Y = 1/2$ (central) and $Y = 2 Y_{\rm max}/3$ (right) for different values of $u$,
    using PDF4LHC21 NNLO PDF set and considering photon-mediated Drell-Yan production at LHC $\sqrt{s}=13$~TeV.}
  \label{fig:lumi}
\end{figure}

To appreciate the size of this effect we plot in figure~\ref{fig:lumi} the ratio of the derivative of the true luminosity over the luminosity in $z=1$
as a function of $\tau$ for different values of $Y=0, Y_{\max}/2, 2Y_{\rm max}/3$ and $u=0,1/2,1$,
with $Y_{\rm max} = \frac12 \log \frac1\tau$,
using the PDF4LHC21 NNLO PDF set~\cite{PDF4LHCWorkingGroup:2022cjn}
and assuming photon-mediated Drell-Yan production at LHC $\sqrt{s}=13$~TeV.
At central rapidity $Y=0$ even with real PDFs we see that the luminosity is almost independent of $u$,
and we observe that for a wide range of $\tau$ values $10^{-4}<\tau<0.1$ this ratio is of $\Ord(1)$,
making the proof of the previous section valid for these kinematics.
We see however a clear growth of this ratio going towards large $\tau$,
confirming that the assumption Eq.~\eqref{eq:lumid} breaks down at some point when $\tau$ is too large.
Increasing the rapidity, the situation gets worse and the ratio becomes larger at smaller values of $\tau$ for some values of $u$
(in this case, having used positive rapidity, the largest effect is at $u=1$).
Despite this deterioration, we notice that in the phenomenologically interesting region of mid-low $\tau$
this ratio remais of $\Ord(1)$ even at large rapidities.

We can thus conclude that the BNX and BFR approaches are formally valid in a restricted kinematical region.
They are not supposed to be accurate at large $\tau$ and towards the rapidity endpoints.
This is somewhat surprising and counterintuitive as these are precisely the regions identified by the hadronic threshold limit,
where threshold resummation is certainly relevant.
The point is that the BNX and BFR formulation, as already discussed, are based on an approximation of the dependence
of $z$ and $u$ of the coefficient function, and it is \emph{this} approximation that is not valid in the hadronic threshold region.
Far from it, the approximation is legitimate, and it allows to resum the partonic threshold logarithms in the cross section.

We can try to estimate in a more quantitative way in which kinematic region
the missing contributions from the derivative of the luminosity make the BNX and BFR formulation break down.
To do so, we consider the BNX and BFR expressions, Eq.~\eqref{eq:BNX_form} and Eq.~\eqref{eq:BFR_form},
in which we replace $C_{\rm thr}(z)$ with the full $C(z)$:
\begin{align}
  \label{eq:BNX_Cexact}
\frac{1}{\tau\sigma_0} \frac{d^2 \sigma^{\rm BNX}_{q \bar{q}} }{d Q^2 \; dY} &\to \int_{\tau}^{1} \frac{dz}z \; \frac{ \Lum_{q \bar{q}}(z,0) + \Lum_{q \bar{q}}(z,1)}{2} C(z),\\
	\label{eq:BFR_Cexact}
\frac{1}{\tau\sigma_0} \frac{d^2 \sigma^{\rm BFR}_{q \bar{q}} }{d Q^2 \; dY} &\to \int_{\tau}^{1} \frac{dz}z \; \Lum_{q \bar{q}}\bigg( z, \frac{1}{2} \bigg) C(z). 
\end{align}
These expressions represent alternative formulations for the rapidity distribution
where only the luminosity is approximated,
and correspond to neglecting the $\Ord(1-z)$ contributions in the penultimate line of Eq.~\eqref{eq:proof}.\footnote
{Of course, these expressions are not suitable for resummation, as the exact $C(z)$ is not known to all orders.}
As such, comparing them with the exact distribution (at fixed order), we are able to judge the quality
of the approximation of the luminosity, and thus the impact of the neglected derivative terms.

\begin{figure}[t]
\centering
\includegraphics[width=\textwidth]{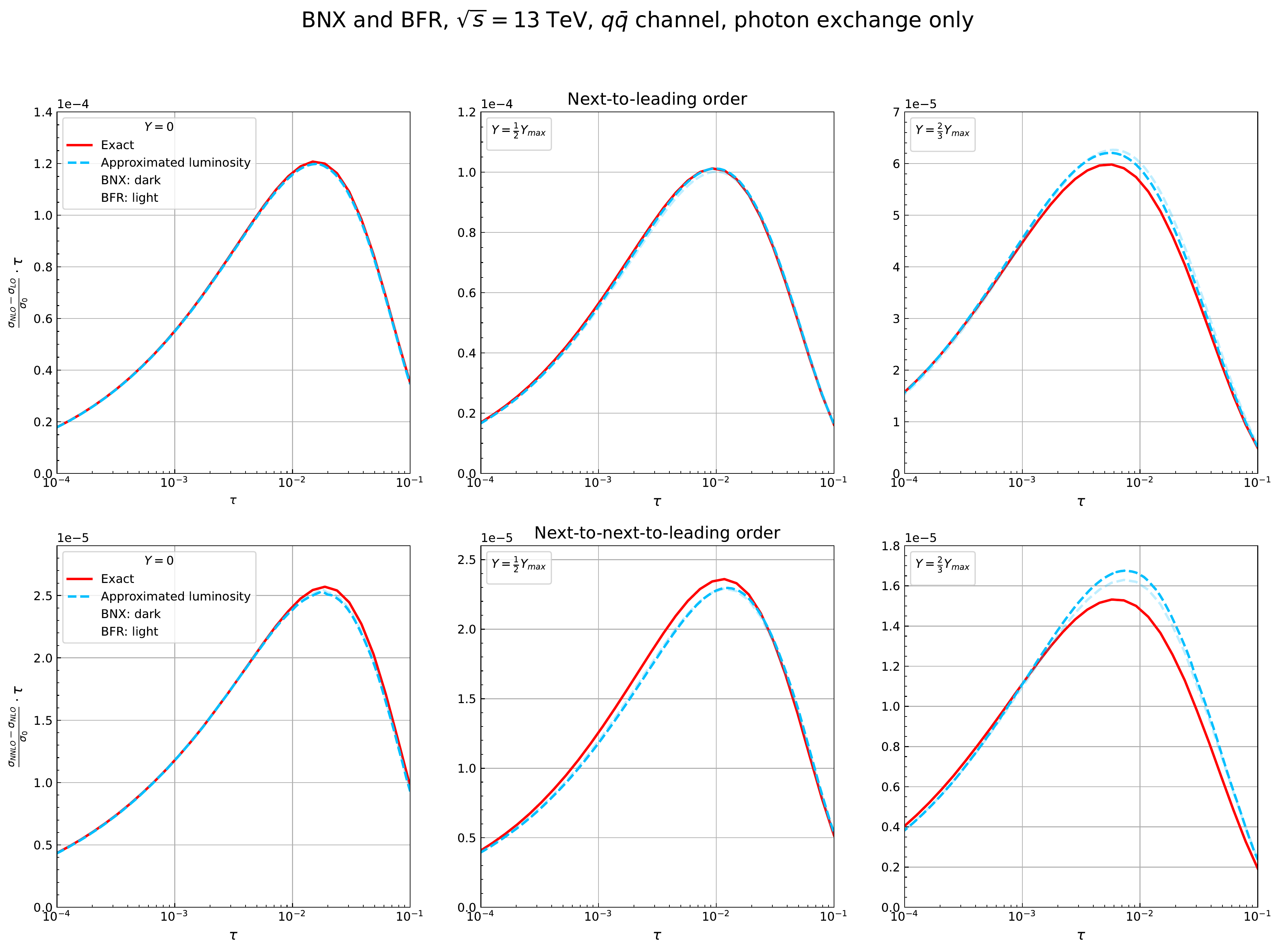}
\caption{Rapidity distributions at NLO (up) and NNLO (down) as a function of $\tau$ for $Y=0,Y_{\rm max}/2,2Y_{\rm max}/3$,
  using PDF4LHC21 NNLO PDF set and considering photon-mediated Drell-Yan production at LHC $\sqrt{s}=13$~TeV.
  The approximations are obtained expanding the luminosity \`a la BNX (darker color) and \`a la BFR (lighter color),
  given by Eq.~\eqref{eq:BNX_Cexact} and \eqref{eq:BFR_Cexact} respectively,
  where we use the full $C(z)$ instead of its threshold approximation $C_{\rm thr}(z)$.}
\label{fig:BNXBFRtauCexact}
\end{figure}

\begin{figure}[t]
\centering
\includegraphics[width=\textwidth]{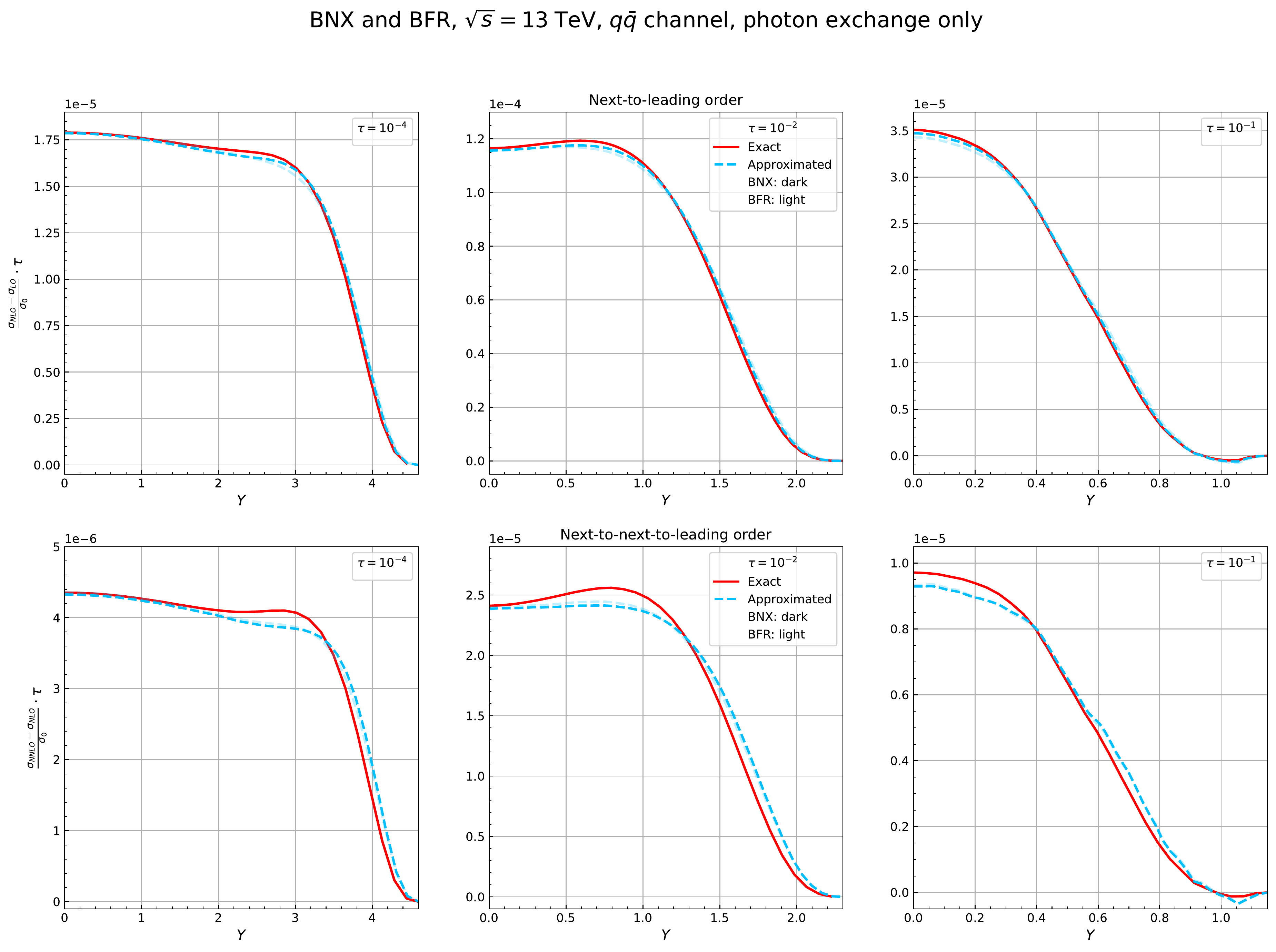}
\caption{Rapidity distributions at NLO (up) and NNLO (down) as a function of $Y$ for $\tau=10^{-4},10^{-2},10^{-1}$,
  using PDF4LHC21 NNLO PDF set and considering photon-mediated Drell-Yan production at LHC $\sqrt{s}=13$~TeV.
  The approximations are obtained expanding the luminosity \`a la BNX (darker color) and \`a la BFR (lighter color),
  given by Eq.~\eqref{eq:BNX_Cexact} and \eqref{eq:BFR_Cexact} respectively,
  where we use the full $C(z)$ instead of its threshold approximation $C_{\rm thr}(z)$.}
\label{fig:BNXBFRrapCexact}
\end{figure}

We do this in figures~\ref{fig:BNXBFRtauCexact} and \ref{fig:BNXBFRrapCexact},
where we plot the exact contributions to the rapidity distribution at NLO and NNLO (in the $q\bar q$ channel only)
along with the approximations of the luminosity dependence \`a la BNX and BFR, Eqs.~\eqref{eq:BNX_Cexact}, \eqref{eq:BFR_Cexact}.
In the first figure the distribution is shown as a function of $\tau$ and for different values of $Y=0, Y_{\max}/2, 2Y_{\rm max}/3$,
while the second figure shows the same distribution but as a function of $Y$ for three values of $\tau=10^{-4}, 10^{-2}, 10^{-1}$.
The plots are obtained considering photon-mediated Drell-Yan production at LHC $\sqrt{s}=13$~TeV,
using again the PDF4LHC21 NNLO PDF set~\cite{PDF4LHCWorkingGroup:2022cjn}
and taking from it the value of the strong coupling.
The exact NNLO result is taken from the \texttt{Vrap} code~\cite{Anastasiou:2003yy,Anastasiou:2003ds},
selecting from it only the terms contributing to the $q\bar q$ channel.

We immediately observe that at small $\tau$ and central rapidity, the quality of the approximation
is excellent at NLO and NNLO, as in both cases the BNX/BFR curves
are almost identical to the exact result.
Note that while we expect a failure of the validity of the expansion in powers of $1-z$ at large $\tau$ and/or large rapidity,
it is perhaps surprising to see such a good agreement at such small values of $\tau$.
Indeed, the neglected $\Ord(1-z)$ contributions, though geniunely subleading, are not
necessarily small as the integration over $z$ extends to values as small as $\tau$.
An explanation of this effect is the fact that the shape of the luminosity strongly favours large values of $z$
in the integrand and suppresses the region of $z$ close to $\tau$,
as a consequence of the small-$x$ growth and the large-$x$ suppression of the PDFs $f_i(x,\muf^2)$, respectively.
Therefore, the integral over $z$ is dominated by the large-$z$ region, well described by a threshold approximation,
while the contribution from medium-small $z$ down to $\tau$ is a small correction.
This phenomenon is the so-called dynamical threshold enhancement of Ref.~\cite{Becher:2007ty},
and it is responsible for the threshold dominance also at the rapidity-integrated level, see e.g.\ Ref.~\cite{Bonvini:2018iwt}.

Moving towards larger values of $Y$ and $\tau$ we see
some deterioration of the agreement, more marked at NNLO.
However, in the range of values of $\tau,Y$ considered here,
the accuracy of the approximation remains very high, with discrepancies of the order of some percent.
We can appreciate in particular a slight distorsion of the $Y$ dependence of the distribution,
clearly visible at NNLO in figure~\ref{fig:BNXBFRrapCexact}.
We stress that, by construction, the integral in rapidity of all the curves in each plot is the same
and coincides with the exact rapidity-integrated cross section, as we also verified numerically.
This constraint also contributes to the high accuracy of the approximation.

We finally notice that the BNX and BFR approaches give basically identical results.
This is a consequence of the fact that the two formulations differ by next-to-next-to-leading power contributions.

The excellent quality of the approximation of the luminosity at the core of the BNX/BFR formulation
can be understood analytically.
By repeating the derivation of Eq.~\eqref{eq:proof} keeping also the linear term in $1-z$ of the expansion
of the luminosity Eq.~\eqref{eq:lumexp}, it is easy to find
\begin{align}
\frac{1}{\tau\sigma_0} \frac{d^2 \sigma_{q \bar{q}} }{d Q^2 \; dY}
&= \int_\tau^{1} \frac{dz}z \[ \Lum_{q \bar{q}}(z,\bar u)\, C(z) + (1-z)\int_0^1du\Big(\Lum'_{q \bar{q}}(1,\bar u)-\Lum'_{q \bar{q}}(1,u)\Big)C(z,u) + \Ord\((1-z)^2\) \]
\end{align}
where the term proportional to $\Lum'_{q \bar{q}}(1,\bar u)$ appears when we replace $\Lum_{q \bar{q}}(1,\cdot)$ with $\Lum_{q \bar{q}}(z,\bar u)$
in the fourth line of Eq.~\eqref{eq:proof}.
We thus observe that the linear term in $1-z$ is not proportional to the entire derivative $\Lum'_{q \bar{q}}(1,u)$,
but to the difference $\Lum'_{q \bar{q}}(1,\bar u)-\Lum'_{q \bar{q}}(1,u)$, which is obviously smaller.
The same holds for higher derivative terms.
In other words, restoring the $z$ dependence in the luminosity, even if this introduces a dependence on
the new, arbitrary variable $\bar u$, is beneficial as it allows to reduce the impact of
the missing contributions at higher order in $1-z$.

We thus conclude that the limitations coming from the growth of the derivative of the luminosity
at large $\tau e^{2|Y|}$ is more formal than practical, and for all phenomenologically
relevant values of these parameters the approximation of the luminosity leading to the BNX/BFR expressions
is fully valid.
In passing, we have also shown that the large-$z$ approximation on the luminosity in Eq.~\eqref{eq:proof}
is of very high quality even when $\tau$ is small,
despite the integral contains many values of $z$ which are far from 1
and for which the threshold expansion is not formally accurate.
The actual quality of threshold resummation based on Eq.~\eqref{eq:proof},
performed either \`a la BNX Eq.~\eqref{eq:BNX_form} or \`a la BFR Eq.~\eqref{eq:BFR_form},
also depends on how well $C_{\rm thr}(z)$ approximates the exact $C(z)$,
and we will discuss this in section~\ref{sec:good}.

\subsection{Objections to BNX and BFR raised in the LMT paper}
\label{sec:leadingpower}

We now analyse the objections raised in Ref.~\cite{Lustermans:2019cau}, which claims that
BNX and BFR are not accurate at leading power in $1-z$.

One of the arguments used in Ref.~\cite{Lustermans:2019cau} is the fact
that a class of terms apparently enhanced in the $z\to1$ limit are missed in these approaches.
Specifically, any term in $C(z,u)$ that vanishes after integration over $u$ cannot be captured by the BNX and BFR approaches.
One such term is for instance the first term appearing in the last line of Eq.~\eqref{eq:C1_NLO} at NLO,
of the form
\begin{align}
\label{eq:term}
F(z,u) = \frac{1}{1-z}\bigg[ \bigg( \frac{1}{u} \bigg)_{+} + \bigg( \frac{1}{1-u }\bigg)_{+} \bigg].
\end{align}
This term is singular in $z=1$, but since it multiplies plus distributions in $u$ it vanishes after integration over $u$.
According to Ref.~\cite{Lustermans:2019cau}, this is a leading power contribution that is missing in BNX and BFR.

We agree with the authors of Ref.~\cite{Lustermans:2019cau} that,
if we were to count powers of $1-z$ at the level of the coefficient function \emph{only},
the term Eq.~\eqref{eq:term} is formally leading power,
like the terms that are retained in BNX/BFR.
However, our derivation of BNX/BFR in section~\ref{sec:proof} adopts a power counting in $1-z$
at the level of the \emph{full} integrand of Eq.~\eqref{eq:XSconv2}, thus including also the parton luminosity.
What we are now going to show is that the term Eq.~\eqref{eq:term}, because of its peculiar $u$ dependence,
contributes at next-to-leading power to the integral,
and is thus consistently missing in the BNX/BFR leading power result.

The proof that the term Eq.~\eqref{eq:term} does not contribute at leading power relies again on the fact that the
luminosity in $z=1$ is independent of $u$.
To see this, we consider the $u$ integral of that term multiplied by the luminosity,
\begin{align}\label{eq:Fbar}
\bar F(z) &\equiv \int_0^1 du\, F(z,u) \Lum_{q\bar q}(z,u) \nonumber\\
&=\frac1{1-z}\(\int_0^1\frac{du}u\, \[\Lum_{q\bar q}(z,u)-\Lum_{q\bar q}(z,0)\] + \int_0^1\frac{du}{1-u}\, \[\Lum_{q\bar q}(z,u)-\Lum_{q\bar q}(z,1)\]  \).
\end{align}
Each difference of luminosities in the integrands can be expanded in powers of $1-z$,
\begin{subequations}\label{eq:Lumexpansions}
\begin{align}
\Lum_{q \bar{q}}(z,u) - \Lum_{q \bar{q}}(z,0)
&= \cancel{\Lum_{q \bar{q}}(1,u) - \Lum_{q \bar{q}}(1,0)} \nonumber \\
&- \bigg( \Lum_{q \bar{q}}^{'}(1,u) - \Lum_{q \bar{q}}^{'}(1,0) \bigg) (1-z)  + \Ord[ (1-z)^2] \\
\Lum_{q \bar{q}}(z,u) - \Lum_{q \bar{q}}(z,1)
&= \cancel{\Lum_{q \bar{q}}(1,u) - \Lum_{q \bar{q}}(1,1)} \nonumber \\
&- \bigg( \Lum_{q \bar{q}}^{'}(1,u) - \Lum_{q \bar{q}}^{'}(1,1) \bigg) (1-z)  + \Ord[ (1-z)^2],
\end{align}
\end{subequations}
and the zeroth-order contribution in each difference vanishes because the luminosity in $z=1$ does not depend on $u$.
Therefore, each integral in the big rounded brackets in Eq.~\eqref{eq:Fbar} is of order $1-z$,
and thus cancels the singularity of the $\frac1{1-z}$ term in front.
We conclude that this contribution behaves as a constant in the $z\to1$ limit,
and therefore counts as a next-to-leading power contribution.
This is also the reason why there is no need to surround this term with a plus distribution,
which would instead be needed if it counted as leading power.

\begin{figure}[t]
  \centering
  \includegraphics[width=\textwidth]{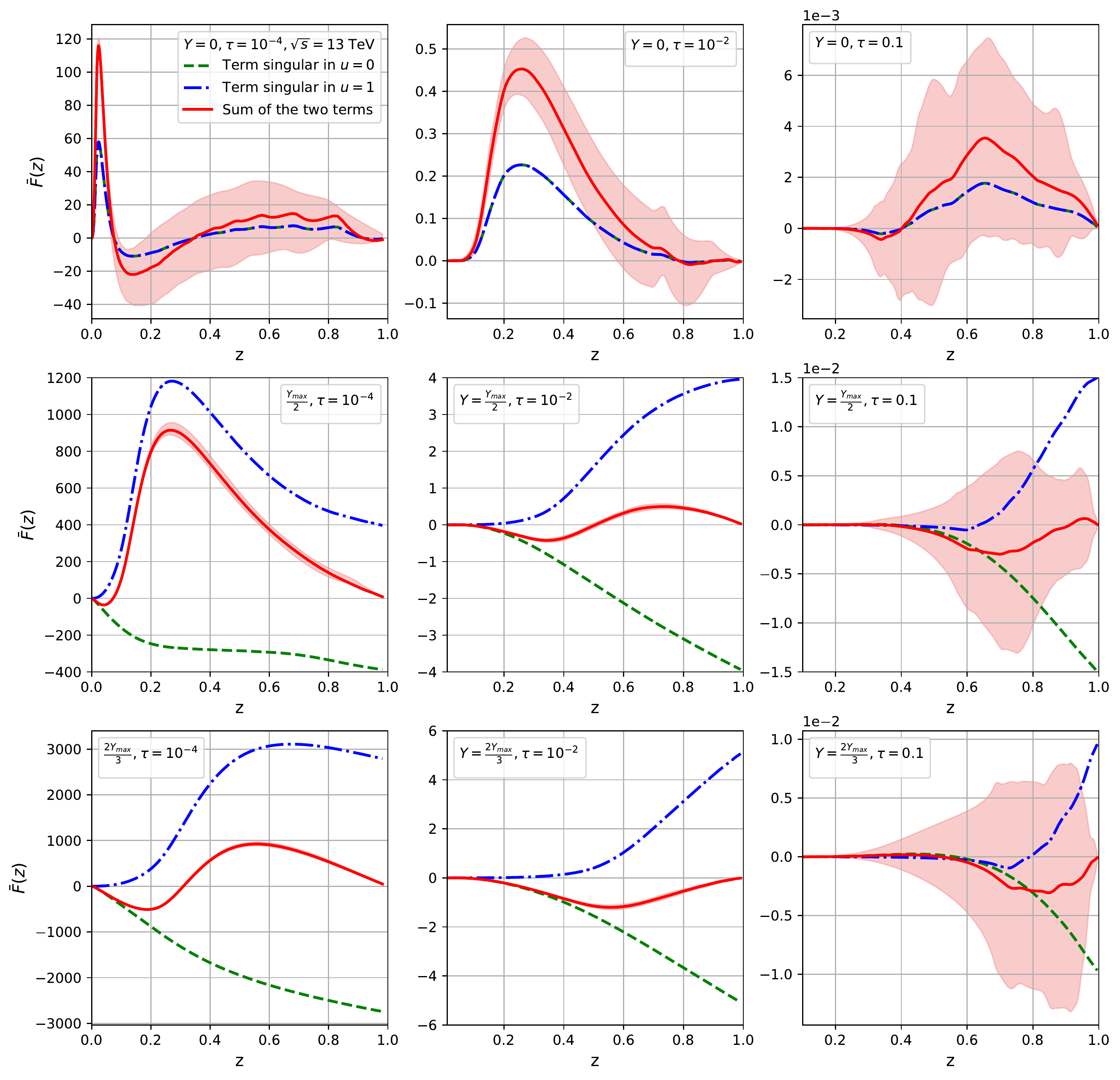}
  \caption{$\bar{F}(z)$ contribution in Eq.~\eqref{eq:Fbar} as a function of $z$ for $\tau = 10^{-4},10^{-2}, 10^{-1}$ and $Y=0,Y_{\rm max}/2,2Y_{\rm max}/3$, using PDF4LHC21 NNLO PDF set and considering photon-mediated Drell-Yan production at LHC $\sqrt{s}=13$~TeV.}
  \label{fig:integrand_uncert}
\end{figure}

We can verify this numerically, by plotting the function $\bar F(z)$ Eq.~\eqref{eq:Fbar} to see that it does not diverge at $z=1$.
We do this in figure~\ref{fig:integrand_uncert}, for different values of $\tau = 10^{-4}, 10^{-2}, 10^{-1}$
and of the rapidity $Y=0, Y_{\max}/2, 2Y_{\rm max}/3$, using the usual physical setup.
Since there are some numerical oscillations at large $z$, probably due to the fact that the curve is the ratio of two small numbers,
we also plot the PDF uncertainty band to make sure that the interpretation of the result is solid.
It is clear that $\bar F(z)$ does not diverge in $z=1$, rather it is perfectly finite,
showing explicitly that this term is next-to-leading power.
We also show (without uncertainty) each of the two contributions to Eq.~\eqref{eq:Fbar}
coming from each integral in the rounded brackets,
as they are separately regular. We see indeed that both of them do not diverge in $z=1$.

In fact, it is possible to note from the plots that the full function $\bar F(z)$ seems to go to zero at $z=1$.
This is indeed the case, as we can verify analytically by noticing that the $\Ord(1-z)$ terms in the expansions Eq.~\eqref{eq:Lumexpansions}
satisfy the relation
\beq
\frac{\Lum_{q \bar{q}}^{'}(1,u) - \Lum_{q \bar{q}}^{'}(1,0)}u + 
\frac{\Lum_{q \bar{q}}^{'}(1,u) - \Lum_{q \bar{q}}^{'}(1,1)}{1-u} = 0,
\eeq
which is easy to prove using the general form of the derivative of the luminosity given by
\beq
\Lum_{q \bar{q}}^{'}(1,u) = -\sum_q c_{q\bar q}
\Big[ u \sqrt\tau e^Y f'_q(\sqrt\tau e^Y) f_{\bar q}(\sqrt\tau e^{-Y})
  + (1-u) \sqrt\tau e^{-Y} f_q(\sqrt\tau e^Y) f'_{\bar q}(\sqrt\tau e^{-Y}) \Big].
\eeq
Therefore, the whole term contributes to the rapidity distribution at next-to-next-to-leading power,
and it is thus more suppressed at threshold than naively expected.
The same is true also for the individual integrals of Eq.~\eqref{eq:Fbar} when $Y=0$, as a consequence
of the fact that the luminosity is symmetric for the exchange $x_1\leftrightarrow x_2$,
that at $Y=0$ corresponds to a symmetry for the exchange $u\leftrightarrow 1-u$.

More in general, the BNX and BFR approaches miss any term in $C(z,u)$ that vanishes after integration in $u$ from 0 to 1.
Consider a generic function $G(z,u)$ that may seem to be leading power
\beq\label{eq:Gdef}
G(z,u) = \frac{\log^k(1-z)}{1-z} g(u),
\eeq
for some integer value of $k$, and with $g(u)$ any function or distribution satisfying the constraint
\beq\label{eq:gu0}
\int_0^1 du\, g(u) = 0.
\eeq
Expanding the luminosity in powers of $1-z$ we find
\begin{align}
  \int_0^1 du\, G(z,u) \Lum_{q\bar q}(z,u)
&= \frac{\log^k(1-z)}{1-z} \int_0^1 du\, g(u) \[\Lum_{q\bar q}(1,\cdot) - \Lum'_{q\bar q}(1,u) (1-z) + \Ord[(1-z)^2]\] \nonumber\\
&= \log^k(1-z)\int_0^1 du\, g(u) \[- \Lum'_{q\bar q}(1,u) + \Ord(1-z)\],
\end{align}
where, thanks to the fact that the luminosity in $z=1$ is $u$ independent,
we could use Eq.~\eqref{eq:gu0} in the last step to show that the first term of the expansion vanishes.
We thus conclude that any apparently leading power term of the general form Eq.~\eqref{eq:Gdef} that vanishes
after integration over $u$ contributes effectively at next-to-leading power.
Therefore, the fact that BNX and BFR miss these contributions does not represent a power-counting issue.

We observe that contributions of this kind are instead present in the approaches to threshold resummation
that keep the separate dependence on $z_a$ and $z_b$,
e.g.\ Refs.~\cite{Banerjee:2018vvb,Catani:1989ne,Mukherjee:2006uu,Ravindran:2006bu,Ravindran:2007sv,Westmark:2017uig,Banerjee:2017cfc}.
These approaches are also supposed to be valid in the threshold $z\to1$ limit,
but since the limit is performed at the level of the coefficient function
these contributions count as leading power and are thus preserved.
For what we have shown, in the strict $z\to1$ limit their inclusion does not increase the accuracy
as they are suppressed with respect to the other leading power terms.\footnote
{Note that this suppression has nothing to do with BNX/BFR:
  it's a general property of the contributions under consideration.}
However, if a threshold approximation/resummation is extended outside the threshold region,
namely for values of $z$ that are not large enough, these contributions may be relevant.
Indeed, a term like Eq.~\eqref{eq:term} contains enhanced contributions in $u\to0,1$,
which are irrelevant at $z\to1$ but not at generic $z$.
To understand this, recall that $u\to0,1$ corresponds to $z_a\to1$ and $z_b\to1$ respectively.
Since $z=z_az_b$, when $z\to1$ both $z_a,z_b\to1$ irrespectively of the value of $u$.
But when $z$ is not large, one among $z_a$ and $z_b$ can be large if $u$ tends to 0 or 1.
These non-threshold but enhanced contributions are captured by the aformentioned approaches,
and improve the description of the tails of the rapidity distribution.
Therefore, the fact that these contributions are missing in BNX and BFR
makes them less accurate than other approaches when the process is far from threshold.

In Ref.~\cite{Lustermans:2019cau} there are other arguments used to criticise the validity of the BNX and BFR approaches.
One of them is an explicit calculation using the same toy PDF
Eq.~\eqref{eq:PDFtoy} showing that after integration in $z$ and $u$ the two functions
\beq
A(z,u) = \plus{\frac1{1-z}}\delta\(u-\frac12\),\qquad
B(z,u) = \plus{\frac1{1-z}}\frac{\delta(u)+\delta(1-u)}2,
\eeq
give rise to different leading power contributions.
As these correspond to a BFR and a BNX implementation of the same term,
if they are both correct at leading power, they should give the same leading power contributions.
The problem here is that the computation of Ref.~\cite{Lustermans:2019cau} assumes $1-x_a\sim1-x_b\ll1$,
namely $\tau\to1$.
We have already commented in section~\ref{sec:limitations} that in this limit the derivation
of BNX and BFR of section~\ref{sec:proof} does not hold anymore, so this conclusion is not surprising.
Moreover, the power counting is performed at hadron level, namely the terms identified in 
Ref.~\cite{Lustermans:2019cau} are leading power in $1-x_a$ and $1-x_b$,
which makes sense because they assume $\tau\to1$ but it cannot be directly related to the leading power terms
in $1-z$ when $\tau$ is not large.

Another objection of Ref.~\cite{Lustermans:2019cau} regarding the BNX and BFR approaches
is related to the expansion of the luminosity Eq.~\eqref{eq:lumexp}.
In particular, LMT say that the expansion of the arguments $x_{1,2}$ of the PDFs,
Eq.~\eqref{eq:x12exp}, to the zeroth order is too trivial because it does not depend on $z$.
More precisely, they say that
``the $u$ dependence is not power suppressed but multiplies the leading dependence of the PDF arguments on $z$ itself,
and so it cannot be dropped''.
We believe that this is not a real issue: the expansion in powers of $1-z$ is legitimate at large $z$,
and if the zeroth order of this expansion is independent of $z$ it cannot represent a reason for expanding to one order higher.
However, for completeness, we can consider an alternative expansion that
overcomes the LMT objection, without affecting the proof Eq.~\eqref{eq:proof}.
Specifically, we can expand in powers of $1-z$ not the full $x_{1,2}$ expression Eq.~\eqref{eq:x12},
but only the square root that depends on $u$:
\begin{subequations}\label{eq:x12exp_v2}
\begin{align}
x_1 &= \sqrt{\frac\tau z}e^Y   \[1+\(u-\frac12\)(1-z) + \Ord((1-z)^2)\], \\
x_2 &= \sqrt{\frac\tau z}e^{-Y} \[1+\(\frac12-u\)(1-z) + \Ord((1-z)^2)\].
\end{align}
\end{subequations}
Now the leading $z$ dependence appears in the zeroth order term of the expansion, which is still $u$ independent.
Of course, this (arbitrary) expansion is equivalent to that of Eq. \eqref{eq:x12exp} up to the order at which it is truncated.
We can write the luminosity as
\beq
\Lum_{q \bar{q}}(z,u) = \Lum^{(0)}_{q \bar{q}}(z) + \Ord(1-z)
\eeq
with
\beq
\Lum^{(0)}_{q \bar{q}}(z) = \sum_q c_{q \bar{q}} f_q\(\sqrt{\frac\tau z}e^Y\) f_{\bar{q}}\(\sqrt{\frac\tau z}e^{-Y}\) = \Lum_{q \bar{q}}\(z,\frac12\),
\eeq
which is again $u$ independent, thus making the rest of the proof identical to what discussed in section~\ref{sec:proof}
(except for the fact that $\Lum^{(0)}_{q \bar{q}}(z)$ can be moved outside the $u$ integral but not the $z$ integral,
but this is immaterial for the derivation of the final result).

A final objection raised in Ref.~\cite{Lustermans:2019cau} is the fact that the original proof of
the BFR approach~\cite{Bonvini:2010tp} has a conceptual problem in one of the steps.
In particular, the proof is based on the expansion of the Fourier transform kernel $e^{iMy}$,
where $M$ is the Fourier conjugate variable to the parton rapidity $y$.
The expansion in powers of $y$ was truncated at order 0 because $y$ is a variable
ranging in $|y|<\frac12\log\frac1z\sim\frac12(1-z)$,
and so higher orders in $y$ are effectively suppressed by powers of $1-z$.
The objection of Ref.~\cite{Lustermans:2019cau} is that the conjugate variable $M$
has to be counted as of order $\frac1{1-z}$ according to the Fourier inversion theorem.
Therefore, the expansion in powers of $y$ is not legitimate, in the sense that the neglected terms are
not really power suppressed.
We agree with this criticism, and confirm that the proof of BFR in Ref.~\cite{Bonvini:2010tp}
is not satisfactory.
Indeed, we also notice that this proof does not make use of the peculiar $z,u$ dependence of the luminosity,
which is instead at the core of the derivation of the BNX and BFR resummation formulas,
as we have shown in section~\ref{sec:proof}.

\section{How good can BNX/BFR be?}
\label{sec:good}

Having established the validity of the BNX and BFR formulation of threshold resummation of rapidity distributions,
we now want to understand how well they approximate the full cross section at fixed order.
This gives us information on how accurate threshold resummation based on BNX or BFR formulations can be.
Despite the fact that there exist resummation formalisms~\cite{Banerjee:2018vvb,Lustermans:2019cau,Ajjath:2021pre}
that are formally superior to BNX and BFR, they remain interesting alternatives for their simplicity,
being based on the resummation of the rapidity-integrated coefficient function which is typically known
for more processes and with higher logarithmic accuracy.

The accuracy of any threshold approximation (and hence resummation) depends on the definition
of the threshold logarithms that are retained. Indeed, any definition of threshold logarithms that differs from
Eq.~\eqref{eq:thrlogs} by subleading power contributions is formally equivalent and thus acceptable,
but the results may differ significantly.
Such a difference may be seen as a limitation of the threshold approximation, as it comes from
sizeable contributions from next-to-leading power contributions
that are beyond the control of leading power threshold resummation.\footnote
{Recently threshold resummation has been extended to next-to-leading
  power~\cite{Ajjath:2021pre,Ajjath:2020ulr,Ajjath:2020lwb,Ajjath:2021lvg,Ajjath:2021bbm,Bhattacharya:2021hae,Ravindran:2022aqr},
  opening up the possibility of pushing the accuracy beyond that of traditional methods.}
However, some subleading power contributions have a universal structure that can be incorporated in the definition
of threshold logarithms, improving the quality of a threshold approximation~\cite
{Kramer:1996iq,Contopanagos:1996nh,Catani:2001ic,Ball:2013bra,Bonvini:2014joa,deFlorian:2014vta,Beneke:2018gvs,Beneke:2019mua}.
In the following we will discuss different choices of threshold logarithms and
compare them numerically against the exact NLO and NNLO results.

\subsection{Definitions of threshold logarithms}
\label{sec:logdef}

Let us focus in this section on the rapidity-integrated coefficient function
\beq
C(z,\as) = \delta(1-z) + \frac{\as}\pi C_1(z) + \(\frac{\as}\pi\)^2 C_2(z) + \Ord(\as^3),
\eeq
which is the ingredient of the BNX and BFR formulations.
The most natural choice for constructing a threshold approximation
is to retain all contributions of the form of Eq.~\eqref{eq:thrlogs} and delta functions.
At order $\as^n$, the coefficient function is thus approximated at threshold as
\beq\label{eq:zsoft}
C_n^{\rm thr}(z) = \sum_{k=0}^{2n-1} c_{n,k} \plus{\frac{\log^k(1-z)}{1-z}} + d_n\delta(1-z),
\eeq
where $c_{n,k}$ and $d_n$ are numerical coefficients (not functions of $z$).
Extending the notation of Refs.~\cite{Ball:2013bra,Bonvini:2014joa},
we shall call the threshold approximation based on Eq.~\eqref{eq:zsoft} $z$-soft approximation,\footnote
{In Ref.~\cite{Ball:2013bra} it was called soft-0.}
meaning the natural threshold approximation in $z$ space.

The $z$-soft approximation is not particularly convenient for two reasons.
One is that it is not very accurate, as we shall see in section~\ref{sec:num}.
The other reason is that it is not easy to construct an all-order resummed result
that contains all and only those contributions.\footnote
{Possible approaches to reproduce the logarithms Eq.~\eqref{eq:zsoft} are the one used in Ref.~\cite{Becher:2007ty}
  based on soft-collinear effective theory and a variant of the one used in Ref.~\cite{Bonvini:2014joa}
  based on the Borel prescription for resummation.}

The most widespread definition of threshold logarithms that is used in resummed
computation is done in Mellin conjugate space, where the phase space of the gluon emissions
(responsible of threshold logarithms) factorises making possible the construction
of an all-order expression in closed form.
Such a definition is based on the expansion at large $N$ (corresponding to the threshold region in Mellin space)
of the Mellin transform of $z$-space logarithmic terms~\cite{Bonvini:2012sh,Bonvini:2014joa}
\beq\label{eq:Nsoft}
\int_0^1 dz\, z^{N-1} \plus{\frac{\log^k(1-z)}{1-z}} = \frac{1}{k+1} \sum_{j=0}^{k+1} {k+1 \choose j}\Gamma^{(j)}(1) \log^{k+1-j}\frac{1}{N}
+\Ord\(\frac1N\),
\eeq
where $\Gamma^{(j)}(x)$ is the $j$-th derivative of the Euler gamma function $\Gamma(x)$.
Neglecting the $\Ord(1/N)$ contributions (which are subleading power at threshold),
this expansion provides an alternative approximation at threshold,
that we call $N$-soft according to the notation of Refs.~\cite{Ball:2013bra,Bonvini:2014joa}.
The $N$-soft approximation can be used directly in $N$ space, which is convenient for instance at resummed level.
It can also be used in the $z$-space formula Eq.~\eqref{eq:zsoft},
by replacing the logarithmic terms with the inverse Mellin transform of the powers of $\log\frac1N$ terms of Eq.~\eqref{eq:Nsoft}.
Details are given in appendix~\ref{sec:approx}.

We also consider an alternative definition of threshold logarithms proposed in Ref.~\cite{Bonvini:2014joa},
whereby $\log N$ is replaced by $\psi_0(N+1)$, where $\psi_0$ is the digamma function.
Indeed, the exact Mellin transform of the threshold logarithms is expressed in terms of polygamma functions $\psi_k(N)$,
all of which go to zero at large $N$ as $N^{-k}$ with the exception of $\psi_0(N)$, which grows as $\log N$.
The argument of $\psi_0(N)$ is further shifted to $N+1$, which is equivalent up to $\Ord(1/N)$.
This choice, denoted $\psi$-soft$_1$ in Ref.~\cite{Bonvini:2014joa}, corresponds to the approximation
obtained by neglecting the $\Ord(1/N)$ contributions in the expansion
\beq\label{eq:psisoft}
\int_0^1 dz\, z^{N-1} \plus{\frac{\log^k(1-z)}{1-z}} = \frac{1}{k+1} \sum_{j=0}^{k+1} {k+1 \choose j}\Gamma^{(j)}(1) \[-\psi_0(N+1)\]^{k+1-j}
+\Ord\(\frac1N\).
\eeq
This approximation can be converted to $z$ space order by order; details are given in appendix~\ref{sec:approx}.
The formal accuracy of this expression is equivalent to that of $N$-soft and $z$-soft,
as they all differ among each other by subleading power contributions at threshold.
However, the $\psi$-soft$_1$ approximation has some advantages that make its quality superior to other choices.

The key observation is that the use of $\psi_0(N+1)$ allows to include at the rapidity-integrated level
subleading power contributions that have a kinematical origin and are thus universal~\cite
{Kramer:1996iq,Contopanagos:1996nh,Catani:2001ic,Ball:2013bra,Bonvini:2014joa,deFlorian:2014vta,Beneke:2018gvs,Beneke:2019mua}.
In particular, the leading logarithmic terms at next-to-leading power in $1-z$ are
predicted correctly to all orders in $\psi$-soft$_1$
in the dominant flavour-diagonal channel for color-singlet production processes such as Higgs and Drell-Yan,
and subleading logarithmic contributions are also partially included.
As a consequence, at the rapidity-integrated level,
the $\psi$-soft$_1$ choice of logarithmic terms
provides better numerical agreement with the exact result than $z$-soft or $N$-soft,
and also leads to a better stabilisation of the scale dependence at resummed level.
These results were obtained for Higgs production~\cite{Ball:2013bra,Bonvini:2014joa},
and they hold also for the Drell-Yan process.
Of course, this simple modification of the resummed logarithms is not able to predict
subleading power contributions coming from the other channels, most importantly the $qg$ channel,
which would be needed for achieving a higher accuracy of the resummation.

Using $\psi$-soft$_1$ for BNX and BFR is expected to improve the resummation of rapidity distributions as well.
Indeed, even though a direct analytical comparison at parton level cannot be done because
$u$ dependence in BNX and BFR is always approximate,
we know that integrating over rapidity Eq.~\eqref{eq:BNX_form} and Eq.~\eqref{eq:BFR_form}
we obtain the correct rapidity-integrated distribution, for which $\psi$-soft$_1$ is known to perform well.
We will now see by a numerical comparison that the $\psi$-soft$_1$ approximation
is more accurate than the others also for rapidity distributions,
thereby providing the most convenient choice of threshold logarithms for an accurate resummation within the BNX/BFR formulation.

\subsection{Numerical validation of BNX/BFR at NLO and NNLO}
\label{sec:num}

In this section we compare the exact NLO and NNLO contributions to the Drell-Yan rapidity distribution
against threshold approximations based on the BNX and BFR formulations and for the three choices of threshold logarithms
discussed in section~\ref{sec:logdef}.

The numerical setup is the same of section~\ref{sec:correct}.
We consider neutral current Drell-Yan production at LHC with $\sqrt s=13$~TeV,
including only the contribution from the photon for simplicity.
We use the PDF4LHC21 NNLO PDF set~\cite{PDF4LHCWorkingGroup:2022cjn},
and take from it the value of the strong coupling.
We always sit at $\muf=\mur=Q$, with $Q=\sqrt{\tau s}$ the invariant mass of the lepton pair.
We only plot the $q\bar q$ contribution to the rapidity distribution.
The exact NNLO result is taken from the \texttt{Vrap} code~\cite{Anastasiou:2003yy,Anastasiou:2003ds}.

\begin{figure}[t]
\centering
\includegraphics[width=\textwidth]{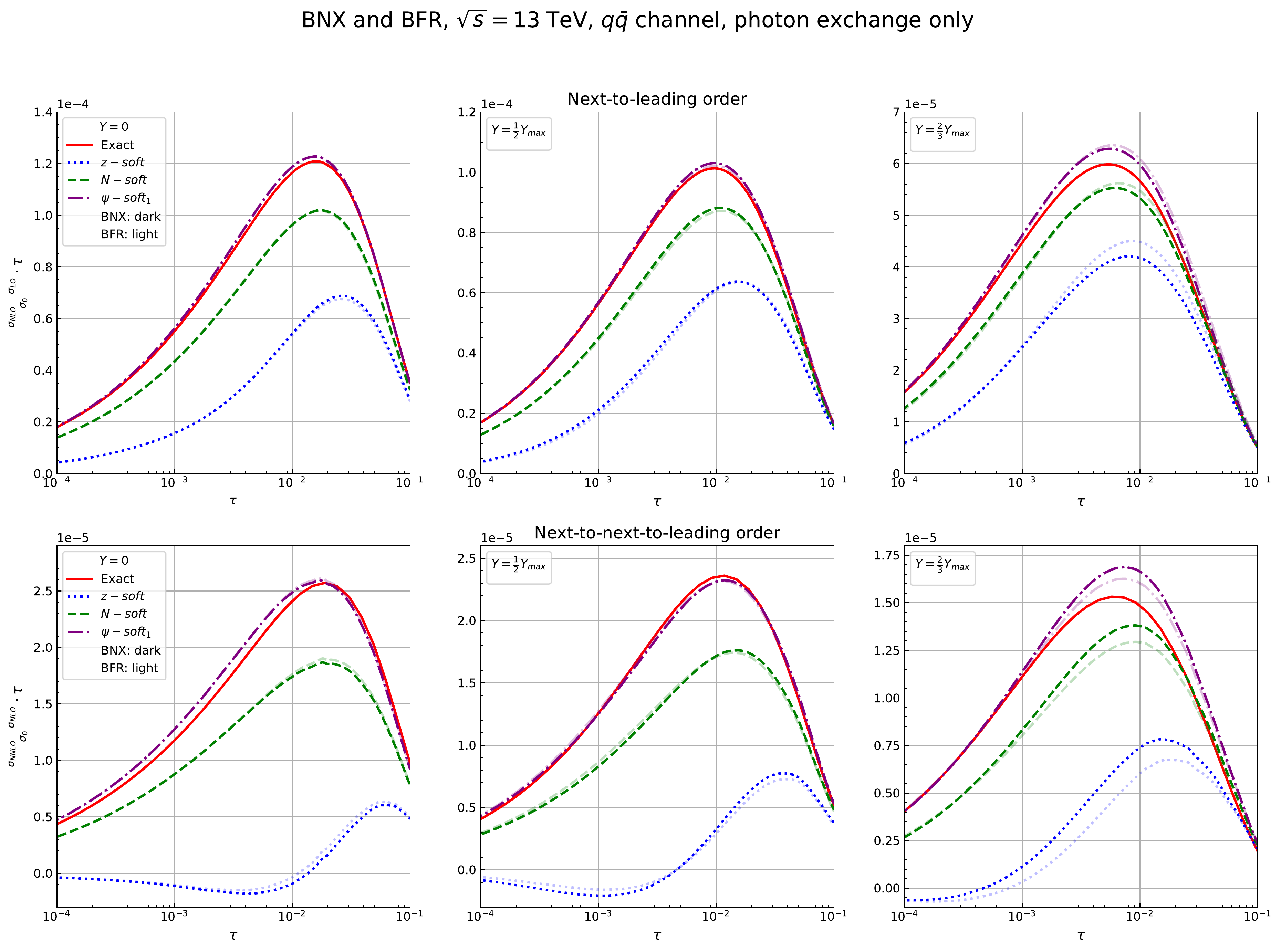}
\caption{Rapidity distributions at NLO (up) and NNLO (down) as a function of $\tau$ for $Y=0,Y_{\rm max}/2,2Y_{\rm max}/3$,
  using PDF4LHC21 NNLO PDF set and considering photon-mediated Drell-Yan production at LHC $\sqrt{s}=13$~TeV.
  The approximations $z$-soft, $N$-soft and $\psi$-soft$_1$ are shown in darker color for BNX and in lighter color for BFR.}
\label{fig:BNXBFRtau}
\end{figure}

We start by showing plots of the rapidity distribution at fixed rapidity and as a function of $\tau$,
i.e.\ as a function of $Q$ since we keep the collider energy fixed.
These are shown in figure~\ref{fig:BNXBFRtau} for BNX (darker curves) and BFR (lighter curves).
In each figure the upper plots show the comparison at pure NLO and the lower plots at pure NNLO.
Each plot corresponds to different values of rapidity, $Y=0,Y_{\rm max}/2, 2 Y_{\rm max}/3$.
The solid red curve represents the exact result, while the other curves are the various approximations
with different definition of threshold logarithms:
dotted blue corresponds to $z$-soft, dashed green is $N$-soft and dot-dashed purple is $\psi$-soft$_1$.

We observe that $\psi$-soft$_1$ is by far the best approximation, in most cases overshooting the exact result
by a small amount.
In fact, comparing with figure~\ref{fig:BNXBFRtauCexact}, we can observe that the $\psi$-soft$_1$ curve
is very similar to the curves obtained with the exact $C(z)$, as a consequence of the aformentioned fact
that $\psi$-soft$_1$ provides an excellent description of the coefficient function at the rapidity-integrated level.
The $N$-soft approximation is reasonably good but definitely worse than $\psi$-soft$_1$,
and it always undershoots the exact by an amount that can reach $35\%$.
Only at large rapidity the large-$\tau$ behaviour seems to be better for $N$-soft than for $\psi$-soft$_1$,
but this is due to an accidental compensation of the deterioration of the approximation of the luminosity
visible in figure~\ref{fig:BNXBFRtauCexact} and the undershooting of $N$-soft.
Finally, $z$-soft is the worst, especially at NNLO where it seems completely unrelated to the exact result,
except at very large $\tau$.
This failure of the $z$-soft approximation is well known and expected at the rapidity-integrated level,
and it has been studied in various works, mostly in the context of Higgs
production~\cite{Catani:1996yz,Catani:2003zt,deFlorian:2012za,Ball:2013bra,Anastasiou:2014lda}.

\begin{figure}[t]
\centering
\includegraphics[width=\textwidth]{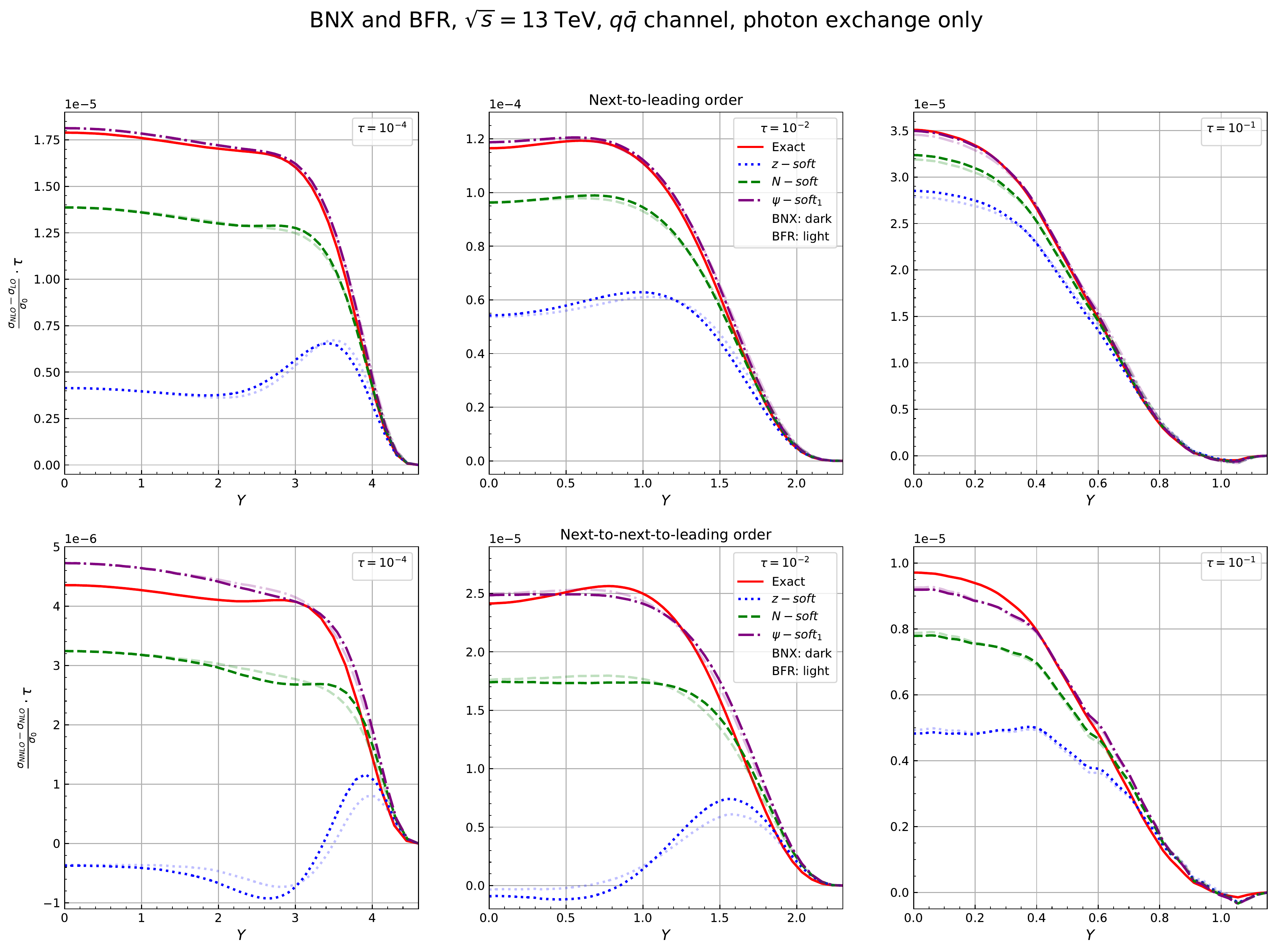}
\caption{Rapidity distributions at NLO (up) and NNLO (down) as a function of $Y$ for $\tau=10^{-4},10^{-2},10^{-1}$,
  using PDF4LHC21 NNLO PDF set and considering photon-mediated Drell-Yan production at LHC $\sqrt{s}=13$~TeV.
  The approximations $z$-soft, $N$-soft and $\psi$-soft$_1$ are shown in darker color for BNX and in lighter color for BFR.}
\label{fig:BNXBFRrap}
\end{figure}

We now move to visualise the same differential distribution as a function of the rapidity $Y$
for fixed values of $\tau$. This is shown in figure~\ref{fig:BNXBFRrap}.\footnote
{At NNLO for $\tau=0.1$ the BNX and BFR results display some oscillations.
  Their presence is not due to numerical instabilities,
  as it depends on the PDF set used, and we verified that with other sets
  they are reduced or go completely away.}
The structure of the plots is the same of the previous figures,
and the three columns correspond to the values $\tau=10^{-4}, 10^{-2}, 10^{-1}$.

We clearly see that at large rapidity the agreement of all the approximations is good,
but moving towards smaller rapidity $z$-soft deviates soon and significantly,
$N$-soft also deviates undershooting the exact result by a large amount,
while $\psi$-soft$_1$ is closer to the exact result,
typically overshooting it by a small amount.
Again, the shape of $\psi$-soft$_1$ is very similar to the result obtained with the full $C(z)$,
  figure~\ref{fig:BNXBFRrapCexact}.
  There is a slight deterioration of the accuracy of $\psi$-soft$_1$ at NNLO with respect to the NLO,
  with the shape at this order being somewhat distorted.
  However, reassuringly, the quality of all the approximations is generally similar at NLO and NNLO,
  showing that the procedure is stable and hopefully preserves its reliability at higher orders.

These consideration hold the same for both BNX and BFR.
While the two approaches give very similar results, in these plots we can see a small difference between them,
which is more marked at medium-small $\tau$ and in the less accurate approximations based on $z$-soft and $N$-soft.
In particular, it seems that BNX is able to better reproduce the little bump
present in the rapidity distribution at the transition
between the central rapidity plateau and the large rapidity drop.
However, the difference between BNX and BFR is so mild and in particular
much smaller than the difference between exact and approximate
(and between the different choices of threshold logarithms)
that it cannot be used to strongly favour one of the two approaches over the other.

In conclusion, we have seen the BNX and BFR approaches,
which can be considered equivalent for all practical purposes,
can describe rather well the exact result, even far from threshold,
provided a good choice of threshold logarithms is used.
The $\psi$-soft$_1$ choice is very convenient as it allows to reach the best
description and at the same time it is very easy to implement at resummed level.
We stress that any ``traditional'' Mellin-space resummation code that uses $N$-soft by default
can be straightforwardly upgraded to $\psi$-soft$_1$ simply replacing $\log N$ with $\psi_0(N+1)$.

\section{Comparison of BNX/BFR with other approaches}
\label{sec:comparison}

Having seen that BNX/BFR are able to approximate sufficiently well the fixed-order result
if a proper choice of the threshold logarithms is made, we now want to compare\footnote
{A numerical comparison of BDDR and BFR was already performed in Ref.~\cite{Banerjee:2018vvb},
  however the choice of logarithms used for BFR was the one of the original paper~\cite{Bonvini:2010tp}
  which differs from the $\psi$-soft$_1$ that we use here.}
these results with the other approaches to threshold resummation of the literature,
namely BDDR~\cite{Banerjee:2018vvb} and LMT~\cite{Lustermans:2019cau}.
We also consider the AMRST~\cite{Ajjath:2021pre} approach, which is the next-to-leading power extension
of BDDR.

For them, we stick to the original definition of threshold logarithms.
For BDDR and AMRST, being them formulated in Mellin space, the logarithms are defined according to $N$-soft,
with two separate logarithms in the variables $z_a,z_b$,
corresponding to logarithms in Mellin space of two distinct variables $N_a$ and $N_b$.
Expressions for BDDR at NLO and NNLO are given in $N_a,N_b$ space in Ref.~\cite{Banerjee:2018vvb}
and can be converted to $z_a,z_b$ space using the results of appendix~\ref{sec:approx}.
For AMRST, we have expanded ourselves the resummed formulas of Ref.~\cite{Ajjath:2021pre}
to order $\as$ and $\as^2$.
Explicit expressions for both BDDR and AMRST at NLO and NNLO are given in appendix~\ref{sec:BDDR},
both in Mellin space and in momentum space.

For LMT, the logarithms used correspond to the $z$-soft definition.
Despite it being a choice that gives inaccurate approximations at leading power,
the fact that LMT includes subleading power contributions partially cures this deficiency.
We stress that the LMT approximation corresponds to the full distributional
part of the coefficient function when written in terms of the $z_a,z_b$ variables.
Explicit expressions at NLO and NNLO are given in appendix~\ref{sec:LMT}.

\begin{figure}[t]
\centering
\includegraphics[width=\textwidth]{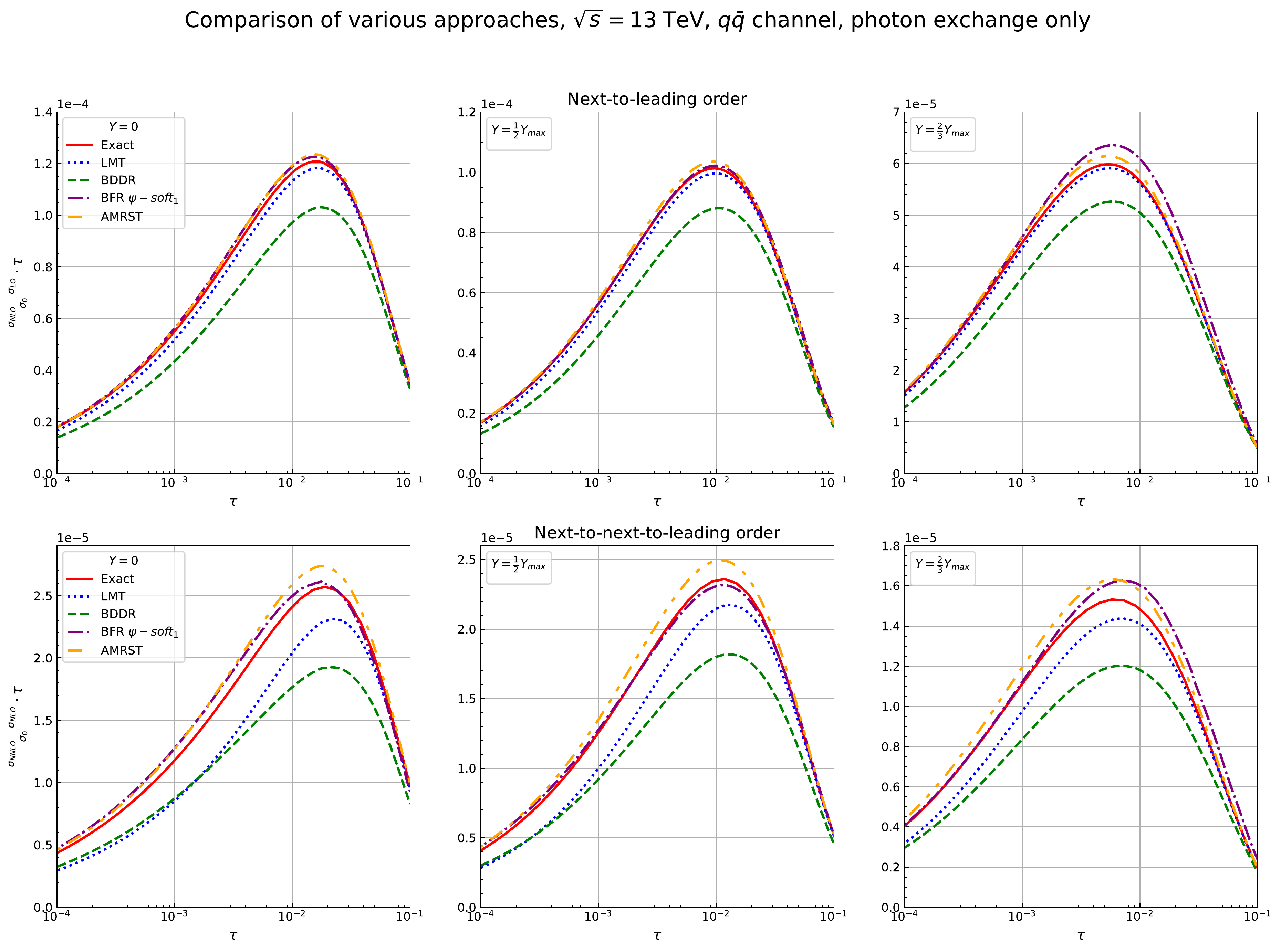}
\caption{Rapidity distributions at NLO (up) and NNLO (down) as a function of $\tau$ for $Y=0,Y_{\rm max}/2,2Y_{\rm max}/3$, comparing the LMT, BDDR, BFR with $\psi$-soft$_1$ and AMRST results and using PDF4LHC21 NNLO PDF set and considering photon-mediated Drell-Yan production at LHC $\sqrt{s}=13$~TeV.}
\label{fig:comparison}
\end{figure}

\begin{figure}[t]
\centering
\includegraphics[width=\textwidth]{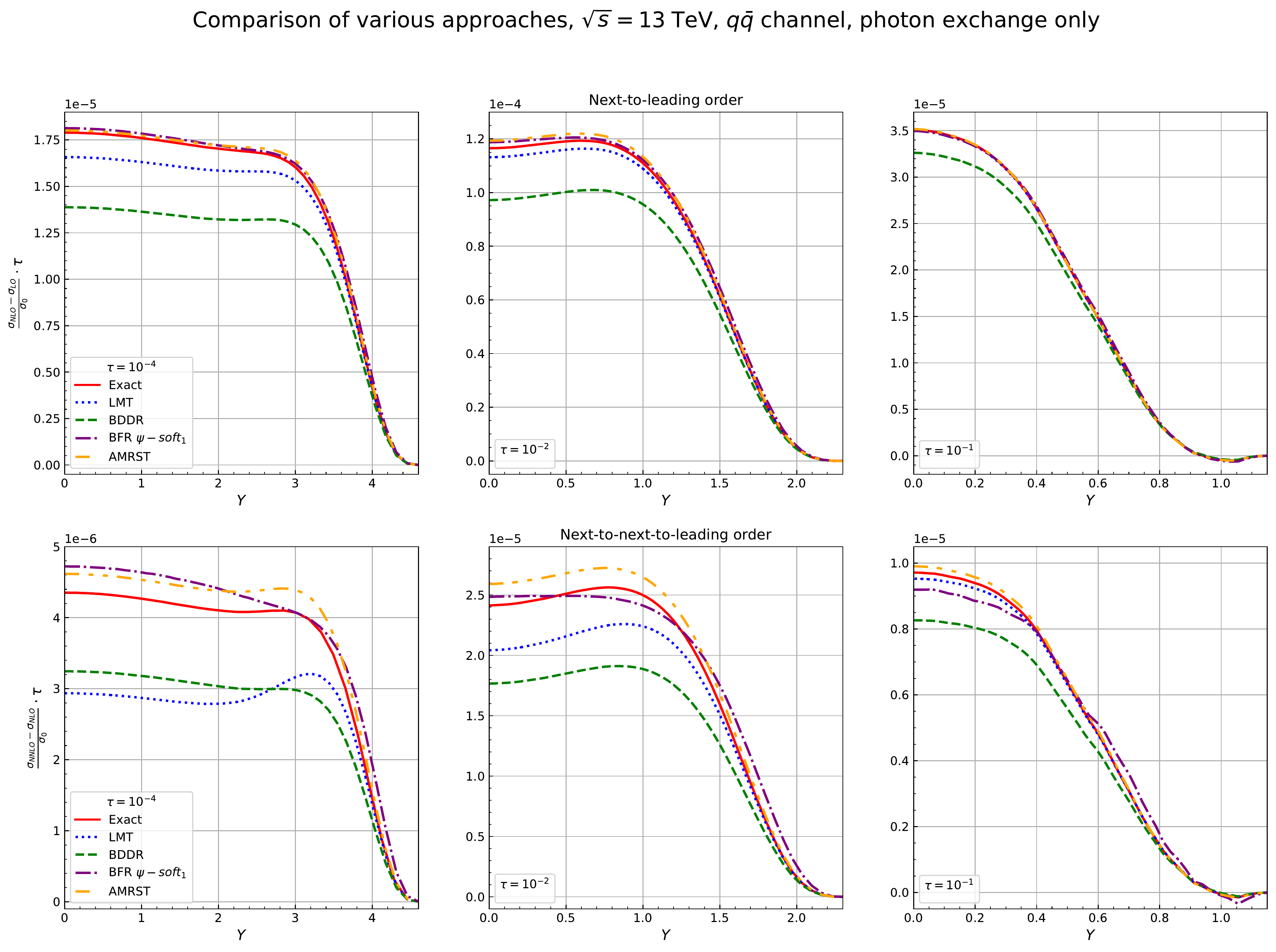}
\caption{Rapidity distributions at NLO (up) and NNLO (down) as a function of $Y$ for $\tau = 10^{-4},10^{-2},10^{-1}$, comparing the LMT, BDDR, BFR with $\psi$-soft$_1$ and AMRST results and using PDF4LHC21 NNLO PDF set and considering photon-mediated Drell-Yan production at LHC $\sqrt{s}=13$~TeV.}
\label{fig:comparisonRap}
\end{figure}

We now compare these results, BDDR, LMT and AMRST, with the BFR approximation based on $\psi$-soft$_1$.
We do not show BNX, that would give results very close to BFR, already presented in section~\ref{sec:good}.
In figure~\ref{fig:comparison} we plot the Drell-Yan rapidity distribution as a function of $\tau$
and in figure~\ref{fig:comparisonRap} as a function of $Y$. The physical setup is the same of the previous plots.
BFR is shown in dot-dashed purple, BDDR in dashed green, LMT in dotted blue and AMRST in dot-dot-dashed orange.

We note that BFR with $\psi$-soft$_1$ is not worse than BDDR,
rather it performs much better both at NLO and NNLO,
being much closer to the exact result.
The reason for this is probably due to the fact that the $N$-soft choice of threshold logarithms of BDDR
is not optimal.
Importantly, this comparison confirms once more that the BNX/BFR formulation of
threshold resummation of rapidity distributions is legitimate and competitive with
canonical, more complex approaches.
Clearly, the shape in rapidity of the BFR curve is only approximate,
and indeed it does not reproduce features of the exact result (e.g., the NNLO bump),
due to the approximate nature of the BFR/BNX approaches.
However, the BFR curve is overall close to the exact result
thanks to the fact that the BFR/BNX approaches are designed to reproduce at the rapidity-integrated level
the inclusive threshold result which is very well approximated by $\psi$-soft$_1$.

Moving on, we observe that,
not surprisingly, the LMT result is generally more accurate than BDDR,
in particular at large rapidity where it is very close to the exact result.
We note however that while at NLO the agreement is very good especially at large $\tau$,
the accuracy of LMT deteriorates significantly at NNLO,
becoming even worse than BDDR at central rapidity for $\tau$ small.
This is certainly a consequence of the use of $z$-soft for the form of the logarithms.\footnote
{This can be also verified numerically by noting that when using a better choice of logarithms the agreement improves,
  as demonstrated by the AMRST result discussed later.}
Indeed, the structure of the LMT result is such that the helpful subleading power
corrections in one variable (say $z_a$)
are retained when they multiply only the leading power terms in the other variable ($z_b$),
which are not accurate enough when using $z$-soft, unless the kinematics forces that variable to be large,
i.e.\ at large rapidity.

We finally move to the AMRST result. Formally, this result contains
less information than the LMT one, as for each variable it only adds the next-to-leading power correction
(multiplied by the leading power term in the other variable),
and not all subleading power corrections as it is done in LMT.
However, the advantage of the AMRST result is that it uses the $N$-soft definition of the threshold logarithms,
which is superior to the $z$-soft one adopted in the LMT approach.
Therefore, as we can see from the figures, the AMRST result is the one that best approximates the exact results.
It is in particular much better than the LMT one at NNLO,
where AMRST stays very close to the exact result, as it also does at NLO.

Notably, BFR with $\psi$-soft$_1$ is closer to the exact result than BDDR and LMT,
with the exception of the high rapidity region.
In fact it is rather close to the AMRST result as well, making BFR (and BNX)
comparable even with the best approach on the market today.
Of course AMRST is superior, as for instance it is able to reproduce the shape
in rapidity which is only vaguely approximated by BFR, as we have already commented.
However,
we believe that BNX/BFR resummation can still prove useful when studying resummation for rapidity distributions.
Indeed, the BFR/BNX approaches are based on the leading-power resummation
of the rapidity integrated cross section, which is available for a large variety of processes
and to a high logarithmic accuracy, while the recent AMRST approach requires
more ingredients and it is only available for a limited number of processes so far.
We will demonstrate the value of the BNX/BFR approaches by showing
representative resummed results in section~\ref{sec:allorder}.

Moreover, motivated by these numerical results,
  we have investigated the analytical difference between the various approaches.
  The details are technical and are collected in appendix~\ref{sec:appcomparison}.

We conclude by stressing that the comparisons presented here are for the $q\bar q$ channel only,
but at next-to-leading power also the other channels contribute.
At the moment only the LMT approach can control the resummation in these subleading channels,
which is a clear advantage for the goal of achieving the highest precision.
It would thus be very interesting to understand if it is possible to modify the LMT approach
by changing the form of the threshold logarithms to take advantage of other better definitions
like $N$-soft or $\psi$-soft$_1$, in order to improve its quality, reaching and possibly surpassing AMRST.
In this way one could achieve the best description of the dominant $q\bar q$ channel,
supplemented by the important contributions from the other subleading channels.
We plan to investigate this possibility in future work.

\section{All-order resummed results for BNX/BFR}
\label{sec:allorder}

Having established the accuracy of the BNX/BFR approaches to resummation,
in this section we present some representative all-order results.
We restrict our attention to BFR resummation (BNX would lead to very similar results),
with the $\psi$-soft$_1$ choice for threshold logarithms.

We perform the resummation using the public
\href{https://www.roma1.infn.it/~bonvini/troll/}{\texttt{TROLL}} code~\cite{Bonvini:2014joa,Bonvini:2015ira,Bonvini:2016frm},
which implements the resummation up to next-to-next-to-next-to-leading logarithmic accuracy (N$^3$LL$^\prime$),\footnote
{The prime notation~\cite{Berger:2010xi,Bonvini:2014qga,Bonvini:2014tea}
  indicates that on top of the purely N$^k$LL contributions
  the constant term (in $N$ space) at N$^k$LO is also included,
  despite it contributing formally at N$^{k+1}$LL in the resummed exponent.
  It is well known that this addition enables the prediction of one extra subleading power of the logarithms
  in the cross section and usually captures most of the next logarithmic order.}
both for rapidity distributions and for rapidity-integrated cross sections.
The ingredients for N$^3$LL$^\prime$ resummation are all available in the literature~\cite{Moch:2005ba,Moch:2005ky,Laenen:2005uz},
including the $\delta(1-z)$ term at N$^3$LO which is taken from Ref.~\cite{Catani:2014uta}
and the recent four-loop cusp anomalous dimension computed in Ref.~\cite{Henn:2019swt}.
We do not exponentiate the constant terms in $N$ space~\cite{Bonvini:2014joa,Bonvini:2016frm},
thus sticking to a more standard attitude, as the effect for this process is rather mild.

\begin{figure}[t]
\centering
\includegraphics[width=\textwidth]{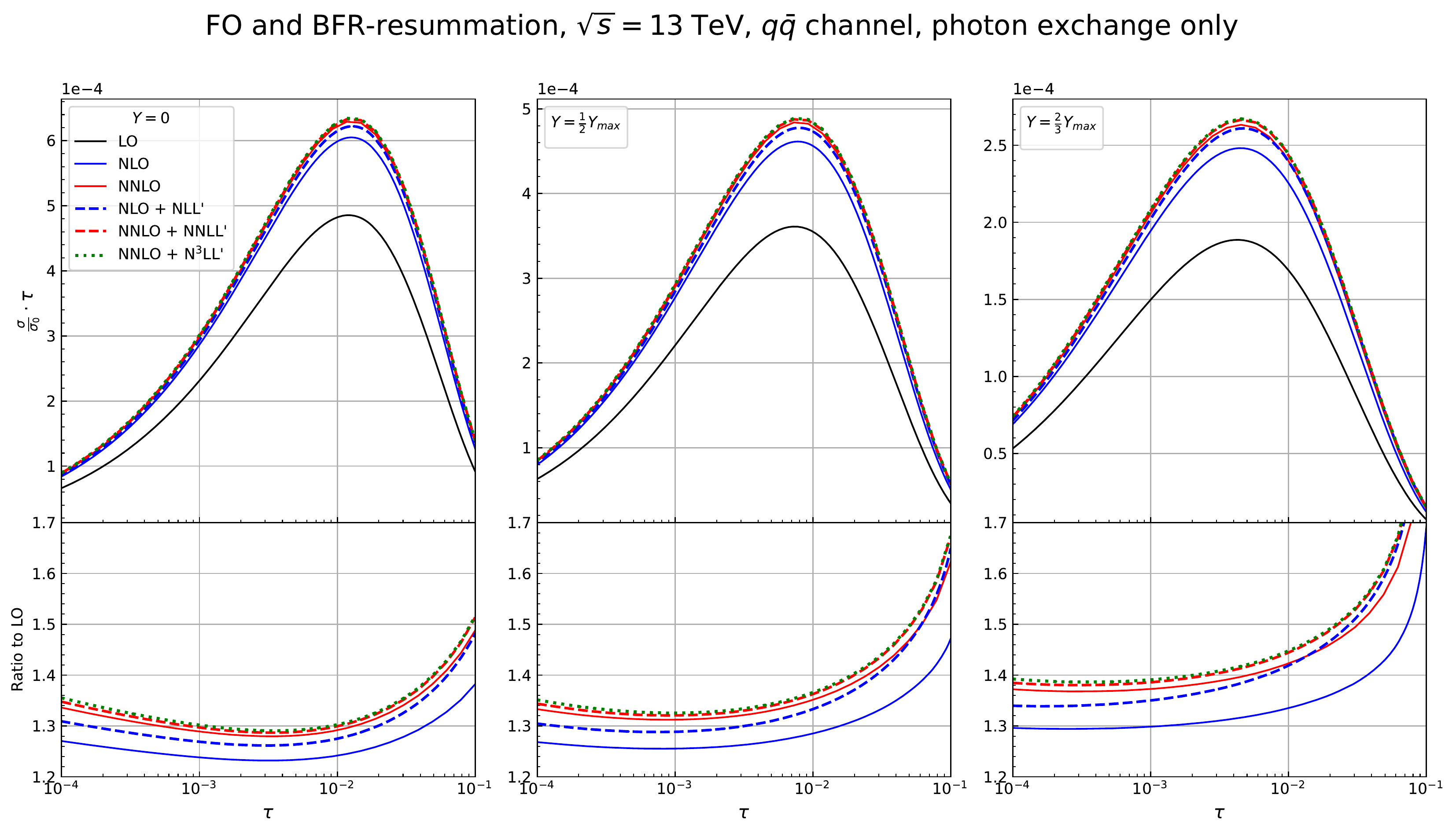}
\caption{Rapidity distributions at fixed order (solid lines) and with resummation \`a la BFR (dashed/dotted lines)
  as a function of $\tau$ for $Y=0,Y_{\rm max}/2,2Y_{\rm max}/3$,
  using PDF4LHC21 NNLO PDF set and considering photon-mediated Drell-Yan production at LHC $\sqrt{s}=13$~TeV.}
\label{fig:BNXBFRtauRes}
\end{figure}

\begin{figure}[t]
\centering
\includegraphics[width=\textwidth]{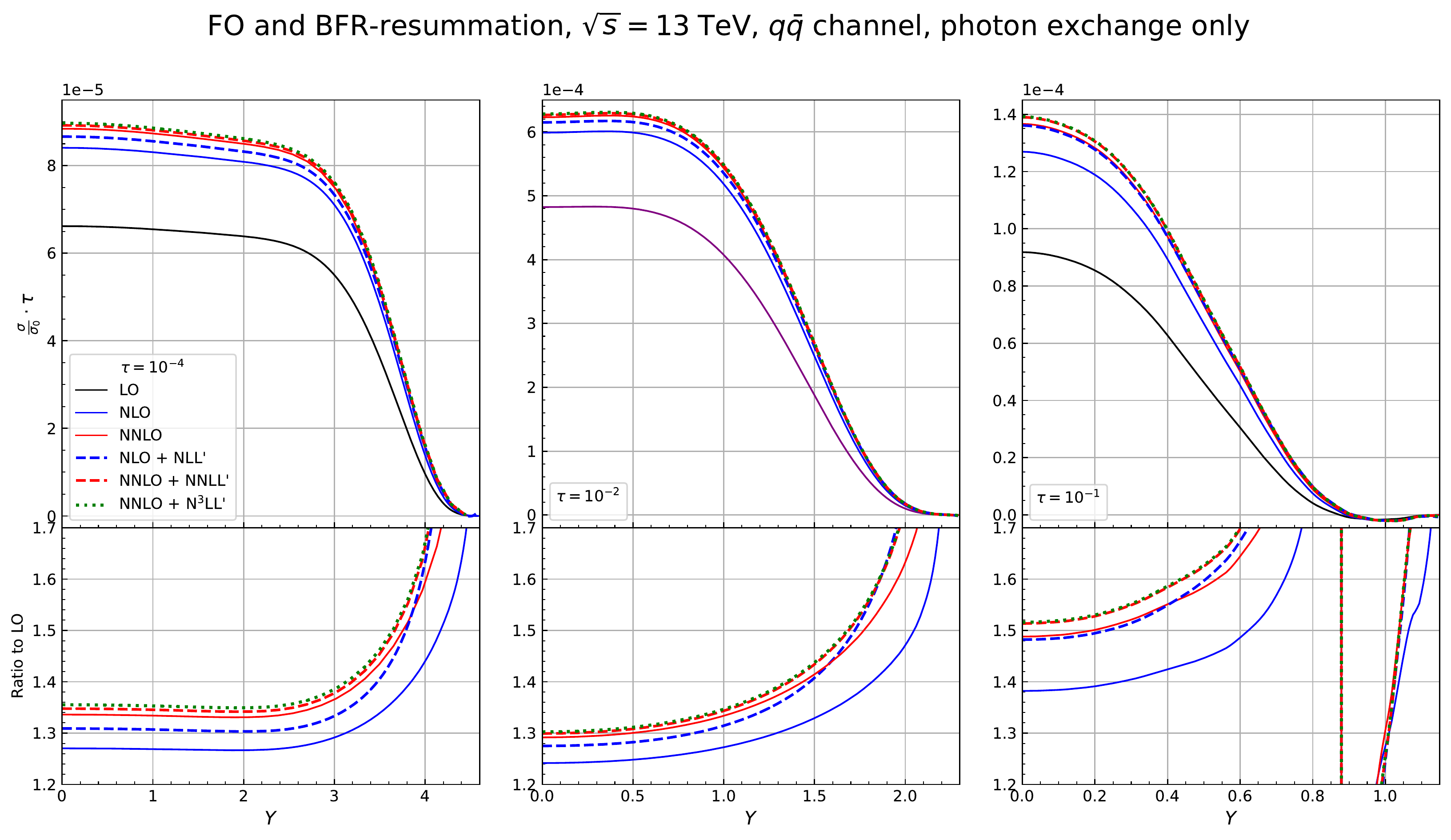}
\caption{Rapidity distributions at fixed order (solid lines) and with resummation \`a la BFR (dashed/dotted lines)
  as a function of $Y$ for $\tau=10^{-4},10^{-2},10^{-1}$,
  using PDF4LHC21 NNLO PDF set and considering photon-mediated Drell-Yan production at LHC $\sqrt{s}=13$~TeV.}
\label{fig:BNXBFRrapRes}
\end{figure}

In figure \ref{fig:BNXBFRtauRes} we show the rapidity distribution as a function of $\tau$ for $Y=0,Y_{\rm max}/2,2Y_{\rm max}/3$
at fixed LO (solid black), NLO (solid blue) and NNLO (solid red) along with the resummed results
at NLO+NLL$^\prime$ (dashed blue),
NNLO+NNLL$^\prime$ (dashed red),
and NNLO+N$^3$LL$^\prime$ (dotted green).
Similarly, in figure~\ref{fig:BNXBFRrapRes} we plot the same curves as a function of $Y$ for $\tau=10^{-4},10^{-2},10^{-1}$.
In all plots we also show a lower panel with the ratio to the LO result, to better appreciate
the relative size of the various perturbative corrections.
As in the previous sections, we only plot the dominant $q\bar q$ channel.

We observe a good convergence of the resummed result, improved with respect to that of the fixed-order result
especially at large $\tau$ and $Y$, where threshold logarithms are more dominant.
Overall, we notice that the effect of adding resummation over the fixed order
is small at NLO and very small at NNLO, which is a consequence of the fact that the Drell-Yan process
exhibits a good perturbative convergence
(as opposed for instance to the Higgs production process in gluon fusion,
see e.g.~\cite{Ahrens:2008qu,Anastasiou:2016cez}).
From the ratio plots,\footnote
{The spike in the third ratio plot of figure~\ref{fig:BNXBFRrapRes} is due to the LO result becoming negative
  at large rapidity.
  This is clearly unphysical, and it is an annoying artefact of the PDFs used that are not very well behaved at large $x$.}
it is apparent that going towards large rapidity or large $\tau$
all the resummed curves tend to overlap, while the fixed-order contributions get larger thus showing a perturbative instability.
This is a consequence of the fact that in these regions the threshold logarithms are dominant and large, 
and resumming them the perturbative expansion stabilizes significantly and leads to reliable perturbatively-stable results.

We were not able to include in the plots the N$^3$LO curve~\cite{Chen:2021vtu} as to our knowledge there is no
public code available out of which we can extract the $q\bar q$ contribution.
Consequently, we could not show the N$^3$LO+N$^3$LL$^\prime$ resummed result,
but we consider the NNLO+N$^3$LL$^\prime$ result, which should be equivalent to the former
in the region of large $\tau$ and $Y$ where the threshold logarithms dominate the distribution.
Far from this region, the addition of the N$^3$LO result would instead improve the accuracy.

In conclusion, we have demonstrated that the BNX/BFR formulations are available
for producing reliable resummed results for rapidity distributions at high logarithmic accuracy.
The public \href{https://www.roma1.infn.it/~bonvini/troll/}{\texttt{TROLL}} code
implements these results up to N$^3$LL$^\prime$ accuracy,
with the $\psi$-soft$_1$ choice of threshold logarithms ($N$-soft is also available).
The results for $\psi$-soft$_1$ resummation for Drell-Yan rapidity distributions are presented here for the first time.

\section{Conclusions}
\label{sec:conclusions}

In this work we have studied threshold resummation of rapidity distributions, applied to the Drell-Yan process.
Our work was motivated by the recent criticism of Ref.~\cite{Lustermans:2019cau}
that states that older approaches to threshold resummation by BNX~\cite{Becher:2007ty} and BFR~\cite{Bonvini:2010tp}
are wrong.

We have proposed a detailed proof of the BNX and BFR approaches, emphasising their limitation
but showing clearly that they are correct within their declared accuracy.
We have rebutted several objections of Ref.~\cite{Lustermans:2019cau},
mostly focussed on the fact that BNX and BFR miss some leading power contributions at threshold,
showing analytically and numerically that this is not the case.

To support our findings we have validated the approaches numerically against the exact NLO and NNLO results.
The quality of these results depends on the choice of the form of threshold logarithms
that one eventually wants to resum to all orders.
We have shown that the $\psi$-soft$_1$ definition of threshold logarithms, proposed in Ref.~\cite{Bonvini:2014joa}
in the context of Higgs production, provides the best results.
We have concluded that despite the approximate nature of the BNX and BFR approaches
they perform rather well as they are able to reproduce to a good accuracy
the exact NLO and NNLO results even far from threshold.

Motivated by this, we have performed a comparison at NLO and NNLO with other approaches in the literature,
from Ref.~\cite{Banerjee:2018vvb} (BDDR), Ref.~\cite{Ajjath:2021pre} (AMRST)
and Ref.~\cite{Lustermans:2019cau} itself (LMT).
We noted that the best approximation is given by the AMRST result,
despite the fact that it is formally less accurate than LMT.
The reason for this is the different form of threshold logarithms used in the two approaches:
the ones used in the LMT paper, namely the standard choice in $z$ space,
have a particularly poor quality in approximating the exact result.
This suggests that the LMT approach could be improved by upgrading the form of threshold logarithms,
and we plan to investigate this in future.

Notably, we have observed that the BNX/BFR approaches with the $\psi$-soft$_1$ definition of threshold logarithms
lead to approximations that are comparable with (if not better than) the others,
despite the latter are all formally more accurate.
This confirms that the BNX/BFR are rather good alternatives to more modern approaches,
and they can still provide a good framework for fast implementations of threshold resummation
in rapidity distributions at high logarithmic accuracy.
We concluded by showing representative results of BFR resummation up to N$^3$LL$^\prime$ accuracy,
obtained through the public \href{https://www.roma1.infn.it/~bonvini/troll/}{\texttt{TROLL}} code,
available at \href{https://www.roma1.infn.it/~bonvini/troll/}{\texttt{http://l.infn.it/mb/troll/}}.

\acknowledgments
{
  We thank J.~Michel and F.~Tackmann for giving us the functions needed to implement the LMT result
  and for discussions.
  We also thank A.~H.~Ajjath, V.~Ravindran and A.~Sankar for correspondence and clarifications on the AMRST result
  and for providing us with some corrections to Ref.~\cite{Ajjath:2021pre}.
  We finally thank S.~Forte and D.~Napoletano for a critical reading of this manuscript.
  The work of G.M.\ is supported by the ERC grant REINVENT No.~714788.
}

\appendix
\section{Analytical expressions}

\subsection{Coefficient function at NLO}
\label{sec:NLO}

We report here the $q\bar q$ coefficient function for rapidity distribution at NLO.
Writing the perturbative expansion of the coefficient function as
\begin{align}
C(z,u,\as)= \delta(1-z) + \frac{\as}{\pi}C_1(z,u) + \(\frac{\as}{\pi}\)^2 C_2(z,u) + \Ord(\as^3),
\end{align}
we can write the NLO coefficient $C_1$ as
\begin{align}
\label{eq:C1_NLO}
  \frac{C_1(z,u)}{C_F} &= \frac{\delta(1-u)+\delta(u)}{2} \bigg[ \(2\zeta_2-4\) \delta(1-z) + 2(1+z^2)\plus{\frac{\log(1-z)}{1-z}}
                         + \log\frac{Q^2}{\mu^2} \plus{\frac{1+z^2}{1-z}} \nonumber \\
                       &\qquad\qquad\qquad\qquad\qquad
                         - \frac{1+z^2}{1-z}\log(z) +1 -z \bigg] \nonumber \\
                       &\quad+ \frac12 \frac{1+z^2}{1-z} \bigg[ \bigg( \frac{1}{u} \bigg)_{+} + \bigg( \frac{1}{1-u} \bigg)_{+} \bigg] - (1-z).
\end{align}
When written in terms of the variables $z_a,z_b$ it becomes
\begin{align}\label{eq:C1tilde}
  \frac{\tilde C_1(z_a,z_b)}{C_F} &= \(3\zeta_2-4+\frac32\log\frac{Q^2}{\mu^2}\) \delta(1-z_a) \delta(1-z_b) 
   \nonumber \\ & \quad
                                    + \plus{\frac{\log(1-z_a)}{1-z_a}} \delta(1-z_b) +  \delta(1-z_a) \plus{\frac{\log(1-z_b)}{1-z_b}}
                                    + \plus{\frac{1}{1-z_a}}\plus{\frac{1}{1-z_b}}
   \nonumber \\ & \quad
                                    + \plus{\frac{1}{1-z_a}}\[\log\frac{Q^2}{\mu^2}\delta(1-z_b)-\frac{1+z_b}2\]
                                    + \[\log\frac{Q^2}{\mu^2}\delta(1-z_a)-\frac{1+z_a}2\]\plus{\frac{1}{1-z_b}}
   \nonumber \\ & \quad
   + \delta(1-z_a) \bigg[ \frac{1-z_b}2 - \frac{1 + z_b}2 \log (1-z_b) +\frac12 \frac{1+z_b^2}{1-z_b} \log\frac{2}{1+z_b}
                  - \frac{1+z_b}2 \log\frac{Q^2}{\mu^2} \bigg]
   \nonumber \\ & \quad
   + \delta(1-z_b) \bigg[ \frac{1-z_a}2 - \frac{1 + z_a}2 \log (1-z_a) +\frac12 \frac{1+z_a^2}{1-z_a} \log\frac{2}{1+z_a}
                  - \frac{1+z_a}2 \log\frac{Q^2}{\mu^2} \bigg]
    \nonumber \\ & \quad
  + \frac{(z_a^2 + z_b^2)[(1+z_a)^2 + (1+z_b)^2 + 2z_az_b(3+z_a+z_b+z_az_b)]}{2(1+z_a)(1+z_b)(z_a + z_b)^2}.
\end{align}

\subsection{Approximations for BNX/BFR}
\label{sec:approx}

In this appendix we provide explicit expressions for the approximations based on the BNX and BFR approaches
at NLO and NNLO used in the text, for all the three choices of threshold logarithms discussed.


In the BNX and BFR approaches the threshold approximation of rapidity distributions is
obtained from the threshold approximation of the rapidity-integrated coefficient function.
At NLO in the $q\bar q$ channel (the only one relevant at threshold) the Drell-Yan coefficient function at threshold
Eq.~\eqref{eq:zsoft} is given by
\beq\label{eq:nlo_thr}
C_1^{\rm thr}(z) = C_F\[4\D_1(z)+2\ell\D_0(z)+\(2\zeta_2-4+\frac32 \ell\)\delta(1-z)\],
\eeq
and at NNLO by
\begin{align}\label{eq:nnlo_thr}
C_2^{\rm thr}(z) &= 8C_F^2 \D_3(z) \nonumber\\
 &+ \[12C_F^2\ell - \frac{11C_A-2n_f}3C_F\]\D_2(z) \nonumber\\
 &+ \[\(4\ell^2+6\ell-16-8\zeta_2\)C_F^2 - \frac{11C_A-2n_f}3C_F\ell +\frac{67C_A-10n_f}9C_F-2\zeta_2C_AC_F\]\D_1(z) \nonumber\\
 &+ \bigg[ \( 3\ell^2 - (8+4 \zeta_2) \ell + 16 \zeta_3 \) C_F^2 
   + \( -\frac{11}{12}\ell^2 + \(\frac{67}{18}-\zeta_2\)\ell + \frac72\zeta_3 + \frac{11}{3}\zeta_2 - \frac{101}{27} \) C_A C_F \nonumber\\
&\quad+ \( \frac1{6} \ell^2 - \frac{5}{9} \ell + \frac{14}{27} - \frac{2}{3} \zeta_2 \) n_fC_F \bigg] \D_0(z) \nonumber\\
 &+ \bigg[ \( \(\frac98 - 2 \zeta_2\)\ell^2 + \(\frac32 \zeta_2 + 11 \zeta_3 - \frac{93}{16}\) \ell
   + \frac1{10}\zeta^2_2 -\frac{35}8 \zeta_2 -\frac{15}4 \zeta_3 + \frac{511}{64} \) C_F^2 \nonumber\\
&\quad+ \( -\frac{11}{16} \ell^2 + \bigg( \frac{193}{48}-\frac32 \zeta_3 \bigg) \ell - \frac{3}{20} \zeta^2_2 + \frac{37}{9} \zeta_2 + \frac74 \zeta_3 - \frac{1535}{192} \) C_A C_F
 \nonumber \\
&\quad+ \( \frac18 \ell^2 -\frac{17}{24}\ell + \frac12 \zeta_3 - \frac{7}{9} \zeta_2 + \frac{127}{96} \) n_f C_F  \bigg]\delta(1-z)
\end{align}
where
\beq
\ell = \lf
\eeq
with $\mu$ the factorization scale, assumed to be equal to the renormalization scale.
In the expressions above we have conveniently written the distributional terms in the compact form defined by
\beq
\D_k(z) \equiv \plus{\frac{\log^k(1-z)}{1-z}}.
\eeq
Using these distributions, Eqs.~\eqref{eq:nlo_thr}, \eqref{eq:nnlo_thr} correspond to the $z$-soft choice of threshold logarithms.


To obtain the $N$-soft expression, we use the results of appendix B.4 of Ref.~\cite{Bonvini:2012sh} to write
\begin{align}\label{eq:Dlog}
\int_0^1 dz\,z^{N-1}\[ \Dl_k(z) + \frac{\Gamma^{(k+1)}(1)}{k+1}\delta(1-z) \] &= \frac{1}{k+1} \sum_{j=0}^{k+1} {k+1 \choose j}\Gamma^{(j)}(1) \log^{k+1-j}\frac1N
\end{align}
where~\cite{Bonvini:2014joa}
\beq \label{eq:DlogDef}
\Dl_k(z) \equiv \plus{\frac{\log^k \log \frac{1}{z}}{\log \frac{1}{z}} }.
\eeq
We recognise in the right-hand side of this equation exactly the dominant large-$N$ limit
of the Mellin transform of the $\D_k(z)$ distribution, Eq.~\eqref{eq:Nsoft}, which is the part retained to build up the $N$-soft
approximation. We immediately conclude that the $N$-soft choice of logarithmic terms corresponds in $z$ space to the replacement
\beq
\D_k(z) \to \Dl_k(z) + \frac{\Gamma^{(k+1)}(1)}{k+1}\delta(1-z).
\eeq
Explicitly, up to $k=3$ as needed for NNLO, we have
\begin{align}
  \D_0(z) &\to \Dl_0(z) - \gamma\delta(1-z), \nonumber \\
  \D_1(z) &\to \Dl_1(z) + \frac{1}{2}\(\gamma^2 + \zeta_2\)\delta(1-z), \nonumber \\
  \D_2(z) &\to \Dl_2(z) - \frac{1}{3}\(\gamma^3 +3 \gamma \zeta_2 +2 \zeta_3\)\delta(1-z),  \nonumber \\
  \D_3(z) &\to \Dl_3(z) + \frac{1}{4}\(\gamma^4 + 6 \gamma^2 \zeta_2 + 8 \gamma \zeta_3 + 3 \zeta_2^2 + 6 \zeta_4\) \delta(1-z),
\end{align}
where $\gamma$ is the Euler-Mascheroni constant.
We also notice that we can write
\beq
\Dl_k(z) + \frac{\Gamma^{(k+1)}(1)}{k+1}\delta(1-z)
= \D_k(z) + \frac{\log^k \log \frac{1}{z}}{\log \frac{1}{z}} - \frac{\log^k(1-z) }{1-z}
\eeq
that provides an alternative, possibly simpler, implementation of $N$-soft.


To obtain an analogous translation rule for $\psi$-soft, let us first consider the
distributions $\Dh_k(z)$ defined by~\cite{Bonvini:2014joa}
\beq\label{eq:Dhat}
\Dh_k(z) = \D_k(z) + \frac{\log^k \frac{1-z}{\sqrt{z} } }{1-z} - \frac{\log^k(1-z) }{1-z}.
\eeq
Morally, this expression is equivalent to including a $\frac1{\sqrt z}$ factor in the argument of the threshold logarithm,
which has a kinematical origin and is thus universal~\cite{Forte:2002ni, Becher:2007ty, Ball:2013bra}.
Practically, this is done in such a way that the Mellin transform of the $\Dh_k(z)$ differ
by that of $\D_k(z)$ by $1/N$ terms, thus providing an equivalent (at leading power) but possibly better
definition of threshold logarithms.

Using again the results of appendix B.4 of Ref.~\cite{Bonvini:2012sh}, we can write the Mellin trasform of the $\Dh_k(z)$ distributions as
\begin{align}
\int_0^1 dz\, z^{N-1}\, \Dh_k(z) = \frac{1}{k+1} \sum_{j=0}^{k+1} {k+1 \choose j} \Gamma^{(j)}(1) \Upsilon_{k+1-j}(N,0)
\end{align}
where
\begin{align}
\Upsilon_0(N,\xi) &= \frac{\Gamma(N-\xi/2)}{\Gamma(N+\xi/2)}
\end{align}
and $\Upsilon_n(N,\xi)$ is the $n$-th derivative of $\Upsilon_0(N,\xi)$ with respect to $\xi$.
By direct computation, we find for the first few terms (as needed for NNLO)
\begin{align}
\Upsilon_1(N,0) &= - \psi_0(N), \nonumber \\
\Upsilon_2(N,0) &= \psi^2_0(N), \nonumber \\
\Upsilon_3(N,0) &= - \psi^3_0(N) -\frac{1}{4} \psi_2(N), \nonumber\\
\Upsilon_4(N,0) &= \psi^4_0(N) +  \psi_2 (N) \psi_0(N),
\end{align}
where $\psi_n(N) = \frac{d^{n+1}}{dN^{n+1}}\log\Gamma(N)$ is the polygamma function.
We observe~\cite{Bonvini:2014joa} that
\beq
\Upsilon_k(N,0) = [-\psi_0(N)]^k \[1+\Ord\(\frac1{N^2}\)\],
\eeq
namely up to next-to-next-to-leading power corrections the use of $\Dh_k(z)$ Eq.~\eqref{eq:Dhat}
can be obtained from a $N$-soft expression Eq.~\eqref{eq:Dlog} with the replacement $\log N\to\psi_0(N)$.
The $\psi$-soft formulation consists in using the distributions $\Dh_k(z)$ for the threshold logarithms
but ignoring these subleading $\Ord(1/N^2)$ contributions
and retaining only the powers of $\psi_0(N)$, which is easy to implement to all orders in $N$ space.
A $z$-space analog would be implemented by a modified distribution $\Dp_k(z)$ defined by
\begin{align}
\int_0^1 dz\, z^{N-1}\, \Dp_k(z) = \frac{1}{k+1} \sum_{j=0}^{k+1} {k+1 \choose j} \Gamma^{(j)}(1) \[-\psi_0(N)\]^{k+1-j},
\end{align}
for which however we cannot find an easy closed form for any $k$.
Up to the order we are interested in, we have that $\Dp_0(z)=\Dh_0(z)$ and $\Dp_1(z)=\Dh_1(z)$, while
\begin{align}
\int_0^1 dz\, z^{N-1}\, \[\Dp_2(z) - \Dh_2(z)\] &=  \frac{\psi_2(N)}{12}, \nonumber\\
\int_0^1 dz\, z^{N-1}\, \[\Dp_3(z) - \Dh_3(z)\] &= - \frac{\psi_2(N)}{4}(\psi_0(N) + \gamma).
\end{align}
The inverse Mellin transforms of these differences can be computed analytically, using e.g.\ the results of Refs.~\cite{Bonvini:2012sh,Blumlein:1998if}.
We find
\begin{align}
  \Dp_2(z) &= \Dh_2(z) - \frac{1}{12} \frac{\log^2(z)}{1-z}, \\
  \Dp_3(z) &= \Dh_3(z)
             - \frac{\zeta_2}2 \frac{\log z}{1-z}+\frac1{12}\frac{\log^3z}{1-z} - \frac12 \frac{\Li_2(z) \log z}{1-z} + \frac{\Li_3(z)}{1-z} - \frac{\zeta_3}{1-z} - \frac{1}{4} \frac{\log(1-z)\log^2z}{1-z}. \nonumber
\end{align}
The $\psi$-soft$_1$ prescription mentioned in the main text uses $N+1$ rather than just $N$
as the argument of the digamma function: $\psi_0(N+1)$.
This is the simplest form of the ``collinear improvement'' introduced in Ref.~\cite{Bonvini:2014joa}
that includes subleading power contributions from universal splitting functions.
In $z$ space a shift $N\to N+1$ corresponds to multiplication by $z$, so the recipe for
the conversion of Eqs.~\eqref{eq:nlo_thr} and \eqref{eq:nnlo_thr} to $\psi$-soft$_1$ is simply given by the replacement
\beq
\D_k(z) \to z\, \Dp_k(z).
\eeq

\subsection{Approximation based on the BDDR/AMRST approach to resummation}
\label{sec:BDDR}

The expansion of the resummed result of BDDR~\cite{Banerjee:2018vvb} (at leading power)
and AMRST~\cite{Ajjath:2021pre} (at next-to-leading power) in $N$ space
is given at NLO by (for $\muf=\mur=Q$)
\begin{align}\label{eq:C1tildeBDDR}
 \tilde C_1(N_a,N_b) &= C_F\bigg[ \frac{\Lb^2}{2}+4\zeta_2-4 + \frac{\Lb}{2\bar N}+\Ord\(\frac1{\bar N^2}\) \bigg]
\end{align}
and at NNLO by\footnote
{We stress that the published version of Ref.~\cite{Ajjath:2021pre} contains some typos that we pointed out to the authors,
  who have fixed them in the latest arXiv version.}
\begin{align}\label{eq:C2tildeBDDR}
  \tilde C_2(N_a,N_b) &= \frac{C_F^2}8\Lb^4 + C_F\frac{11C_A-2n_f}{72} \Lb^3 
                        +\[C_F^2(2\zeta_2-2)+C_FC_A\(\frac{67}{72}-\frac{\zeta_2}4\)-C_F n_f\frac5{36}\]\Lb^2 \nonumber\\
                      &\quad +\[C_F C_A\(\frac{101}{54}-\frac74\zeta_3\)-C_F n_f\frac7{27} \]\Lb
                        +C_F^2\(\frac{511}{64}-\frac{99}8\zeta_2+\frac{69}{10}\zeta_2^2-\frac{15}4\zeta_3\)\nonumber\\
                      &\quad + C_F C_A \(-\frac{1535}{192}+\frac{47}6\zeta_2-\frac{23}{20}\zeta_2^2+\frac{151}{36}\zeta_3\)
                        +C_F n_f \(\frac{127}{96}-\frac43\zeta_2+\frac{\zeta_3}{18}\) \nonumber\\
                      &\quad + \frac{C_F^2}4 \frac{\Lb^3}{\bar N}
                        + \[\frac{C_F^2}8\frac1{\bar N}+C_F\frac{11C_A-2n_f}{48}\]\frac{\Lb^2}{\bar N}
                        +C_F^2\(\frac{\Lb_a}{N_a} + \frac{\Lb_b}{N_b}\)\Lb
                        -\frac{C_F^2}4\(\frac{\Lb_a^2}{N_a} + \frac{\Lb_b^2}{N_b}\)\nonumber\\
                      &\quad + \[C_F^2\(2\zeta_2-\frac{11}4\)
                        +C_FC_A\(\frac{133}{72}-\frac{\zeta_2}4\)
                        -C_Fn_f\frac{11}{36}\]\frac{\Lb}{\bar N}\nonumber\\
                      &\quad + \[ C_F^2\frac58-C_FC_A\frac58 \]\(\frac{\Lb_a}{N_a} + \frac{\Lb_b}{N_b}\) \nonumber\\
                      &\quad + \bigg[-C_F^2\frac{\zeta_2}4
                        -C_Fn_f\frac{19}{54}
                        +C_FC_A\(\frac{97}{216}-\frac78\zeta_3\)\bigg]\frac1{\bar N}  +\Ord\(\frac1{\bar N^2}\)
\end{align}
with
\begin{align}\label{eq:LbarNbar}
  \frac1{\bar N} &\equiv \frac1{N_a} + \frac1{N_b} &
  \Lb_{a,b} &\equiv \log N_{a,b} + \gamma &
  \Lb &\equiv \log(N_aN_b)+2\gamma = \Lb_a+\Lb_b.
\end{align}
To convert these results to $z$ space, we can invert Eq.~\eqref{eq:Dlog} for the leading power terms,
while for the next-to leading power contributions we further need the relation
\beq
\int_0^1 dz\, z^{N-1}\, \log^p\frac1z\, \log^k\log\frac1z = \frac1{N^{1+p}} \sum_{j=0}^k\binom{k}{j}\Gamma^{(j)}(1+p)\log^{k-j}\frac1N,
\qquad k,p\geq0,
\eeq
which can be derived from the generating function $\log^\xi\frac1z$ deriving $k$ times with respect to $\xi$ in $\xi=p$.
Using these results we get at NLO
\begin{align}
 \tilde C_1(z_a,z_b) &= C_F\bigg[ \(\Dl_1(z_a) -\gamma\Dl_0(z_a)\)\delta(1-z_b) + \delta(1-z_a) \(\Dl_1(z_b) -\gamma\Dl_0(z_b)\)  \nonumber\\
                     &\qquad\quad + \Dl_0(z_a) \Dl_0(z_b)+ \(4\zeta_2-4+2\gamma^2\) \delta(1-z_a) \delta(1-z_b) \bigg] \nonumber\\
&\quad +\frac{C_F}{2} \[\(\gamma-\log\log\frac1{z_b}\) \delta(1-z_a)+\(\gamma-\log\log\frac1{z_a}\) \delta(1-z_b) -\Dl_0(z_a)-\Dl_0(z_b)\]
\end{align}
and at NNLO
\begin{align}
  \tilde C_2(z_a,z_b) &= \bigg[C_F^2\(\frac{511}{64}-\frac{99}8\zeta_2+\frac{69}{10}\zeta_2^2-\frac{15}4\zeta_3+8\zeta_2\gamma^2-8\gamma^2+2\gamma^4\)
                        \nonumber \\ & \qquad
                        +C_FC_A\(-\frac{1535}{192}+\frac{47}6\zeta_2-\frac{23}{20}\zeta_2^2+\frac{151}{36}\zeta_3-\zeta_2\gamma^2-\frac72\zeta_3\gamma+\frac{101}{27}\gamma+\frac{67}{18}\gamma^2+\frac{11}9\gamma^3\)
                        \nonumber \\ & \qquad
                        +C_Fn_f \(\frac{127}{96}-\frac43\zeta_2+\frac{\zeta_3}{18}-\frac{14}{27}\gamma-\frac59\gamma^2-\frac29\gamma^3\)\bigg] \delta(1-z_a) \delta(1-z_b) 
   \nonumber \\ & \quad
   + \frac{C_F^2}2 \Big[\Dl_3(z_a) \delta(1-z_b) + \delta(1-z_a) \Dl_3(z_b)\Big]
   \nonumber \\ & \quad
   + \frac32C_F^2 \Big[\Dl_2(z_a) \Dl_0(z_b) + \Dl_0(z_a) \Dl_2(z_b)\Big]
   \nonumber \\ & \quad
   + 3C_F^2 \Dl_1(z_a)\Dl_1(z_b)
   \nonumber \\ & \quad
   +\[-\frac32C_F^2\gamma- C_F\frac{11C_A-2n_f}{24}\] \Big[\Dl_2(z_a) \delta(1-z_b) + \delta(1-z_a) \Dl_2(z_b)\Big]
   \nonumber \\ & \quad
   - C_F\frac{11C_A-2n_f}{12} \Big[\Dl_1(z_a) \Dl_0(z_b) + \Dl_0(z_a) \Dl_1(z_b)\Big]
   \nonumber \\ & \quad
   + \[C_F^2\(\frac52\zeta_2-4+\frac32\gamma^2\)+C_FC_A\(\frac{67}{36}-\frac{\zeta_2}2+\frac{11}{12}\gamma\) +C_Fn_f\(-\frac5{18}-\frac\gamma6\) \]
   \nonumber \\ & \qquad
                  \times\Big[\Dl_1(z_a) \delta(1-z_b) + \delta(1-z_a) \Dl_1(z_b) \Big]
   \nonumber \\ & \quad
   + \[C_F^2\(\zeta_2-4\)+C_FC_A\(\frac{67}{36}-\frac{\zeta_2}2\) -\frac5{18}C_Fn_f\]
                  \Dl_0(z_a)\Dl_0(z_b)
   \nonumber \\ & \quad
   + \bigg[C_F^2\(\zeta_3-\frac52\zeta_2\gamma+4\gamma-\frac{\gamma^3}2\)+C_FC_A\(\frac{11}{24}\zeta_2+\frac74\zeta_3-\frac{101}{54}+\frac\gamma2\zeta_2-\frac{11}{24}\gamma^2-\frac{67}{36}\gamma\)
   \nonumber \\ & \qquad\quad
   +C_Fn_f \(\frac7{27}-\frac{\zeta_2}{12}+\frac{\gamma^2}{12}+\frac5{18}\gamma\)\bigg]
                  \Big[\Dl_0(z_a) \delta(1-z_b) + \delta(1-z_a) \Dl_0(z_b) \Big]
   \nonumber \\ & \quad
   +\bigg\{ \Dl_2(z_a) \bigg[ -\frac34C_F^2\bigg]
   \nonumber \\ & \qquad\quad
   + \Dl_1(z_a) \bigg[-\frac32C_F^2\log\log\frac1{z_b} +C_F\frac{11C_A-2n_f}{24} + \frac{C_F^2}4\log\frac1{z_b} \bigg]
   \nonumber \\ & \qquad\quad
   + \Dl_0(z_a)\bigg[ -\frac34C_F^2\log^2\log\frac1{z_b}+\(C_F^2+C_F\frac{11C_A-2n_f}{24}\)\log\log\frac1{z_b}
   \nonumber \\ & \qquad\qquad\qquad\qquad
                  +C_F^2\(\frac{11}4-\frac{\zeta_2}2\) +C_FC_A\(-\frac{133}{72}+\frac{\zeta_2}4\)+C_Fn_f\frac{11}{36}
   \nonumber \\ & \qquad\qquad\qquad\qquad
                  +\frac{C_F^2}4\log\frac1{z_b}\(\log\log\frac1{z_b}-1\)
                  \bigg]
   \nonumber \\ & \qquad\quad
   + \delta(1-z_a) \bigg[ -\frac14C_F^2\log^3\log\frac1{z_b}+ \(\frac34C_F^2(1+\gamma)+C_F\frac{11C_A-2n_f}{48}\)\log^2\log\frac1{z_b}
   \nonumber \\ & \qquad\qquad\qquad\qquad
                  + \bigg(C_F^2\(\frac{17}8-\frac54\zeta_2-\gamma-\frac34\gamma^2\) +C_FC_A\(-\frac{11}{9}+\frac{\zeta_2}4-\frac{11}{24}\gamma\)
   \nonumber \\ & \qquad\qquad\qquad\qquad\qquad
                  +C_Fn_f\(\frac{11}{36}+\frac\gamma{12}\)\bigg)\log\log\frac1{z_b}
   \nonumber \\ & \qquad\qquad\qquad\qquad
                  +C_F^2\(-\zeta_2-\frac12\zeta_3+\frac54\zeta_2\gamma-\frac{11}4\gamma+\frac14\gamma^3\)
   \nonumber \\ & \qquad\qquad\qquad\qquad
                  +C_FC_A\(\frac{97}{216}-\frac{11}{48}\zeta_2-\frac78\zeta_3-\frac{\zeta_2}4\gamma+\frac{133}{72}\gamma+\frac{11}{48}\gamma^2\)
   \nonumber \\ & \qquad\qquad\qquad\qquad
                  +C_Fn_f\(-\frac{19}{54}+\frac{\zeta_2}{24}-\frac{11}{36}\gamma-\frac{\gamma^2}{24}\)
   \nonumber \\ & \qquad\qquad\qquad\qquad
                  + \frac{C_F^2}8\log\frac1{z_b}\(\log^2\log\frac1{z_b}-2\log\log\frac1{z_b}+2-\zeta_2+\gamma^2\)
                  \bigg]
   \nonumber \\ & \qquad\quad
   +(z_a\leftrightarrow z_b) \bigg\}
   \nonumber \\ & \quad
                  +\frac{C_F^2}4\[\(\log\log\frac1{z_a}+\log\log\frac1{z_b}\)^2-2\zeta_2 \],
\end{align}
having used the $\Dl_k(z)$ distributions defined in Eq.~\eqref{eq:DlogDef}.

\subsection{Approximation based on the LMT approach to resummation}
\label{sec:LMT}

In the generalized threshold expansion of LMT, the expansion of the resummed result can be obtained
from the expression
\beq
\tilde C_{ij}^{\rm LMT}(z_a,z_b,\as) = H_{kr}(\as)\[\delta_{ki}\hat{\cal I}_{rj}(z_a,z_b,\as) + \hat{\cal I}_{ki}(z_b,z_a,\as) \delta_{rj} -\hat S(z_a,z_b,\as) \],
\eeq
in terms of the functions defined in the LMT paper~\cite{Lustermans:2019cau}.
Focussing on the $q\bar q$ channel and expanding in powers of $\as$,
we obtain at NLO at central scales
\begin{align}\label{eq:C1tildeLMT}
  \frac{\tilde C_1^{\rm LMT}(z_a,z_b)}{C_F} &= \big[3\zeta_2-4\big] \delta(1-z_a) \delta(1-z_b) 
                                    + \D_1(z_a) \delta(1-z_b) +  \delta(1-z_a) \D_1(z_b) + \D_0(z_a) \D_0(z_b)
   \nonumber \\ & \quad
                  + \bigg\{
   \delta(1-z_a) \bigg[ \frac{1-z_b}2 - \frac{1 + z_b}2 \log (1-z_b) +\frac12 \frac{1+z_b^2}{1-z_b} \log\frac{2}{1+z_b} \bigg]
                  -\D_0(z_a)\frac{1+z_b}2
                  \nonumber \\ & \qquad\quad
   + (z_a\leftrightarrow z_b)\bigg\},
\end{align}
corresponding to the distributional part of Eq.~\eqref{eq:C1tilde}, and at NNLO we get
\begin{align}\label{eq:C2tildeLMT}
  \tilde C_2^{\rm LMT}(z_a,z_b) &= \bigg[C_F^2\(\frac{511}{64}-\frac{67}8\zeta_2+\frac{19}5\zeta_2^2-\frac{15}4\zeta_3\)
                        +C_FC_A\(-\frac{1535}{192}+\frac{215}{36}\zeta_2-\frac{13}{20}\zeta_2^2+\frac{43}{12}\zeta_3\)
                        \nonumber \\ & \qquad
                        +C_Fn_f \(\frac{127}{96}-\frac{19}{18}\zeta_2+\frac{\zeta_3}6\)\bigg] \delta(1-z_a) \delta(1-z_b) 
   \nonumber \\ & \quad
   + \frac{C_F^2}2 \Big[\D_3(z_a) \delta(1-z_b) + \delta(1-z_a) \D_3(z_b)\Big]
   \nonumber \\ & \quad
   + \frac32C_F^2 \Big[\D_2(z_a) \D_0(z_b) + \D_0(z_a) \D_2(z_b)\Big]
   \nonumber \\ & \quad
   + 3C_F^2 \D_1(z_a) \D_1(z_b)
   \nonumber \\ & \quad
   - C_F\frac{11C_A-2n_f}{24} \Big[\D_2(z_a) \delta(1-z_b) + \delta(1-z_a) \D_2(z_b)\Big]
   \nonumber \\ & \quad
   - C_F\frac{11C_A-2n_f}{12} \Big[\D_1(z_a) \D_0(z_b) + \D_0(z_a) \D_1(z_b)\Big]
   \nonumber \\ & \quad
   + \[C_F^2\(\zeta_2-4\)+C_FC_A\(\frac{67}{36}-\frac{\zeta_2}2\) -\frac5{18}C_Fn_f \]
                  \Big[\D_1(z_a) \delta(1-z_b) + \delta(1-z_a) \D_1(z_b)\Big]
   \nonumber \\ & \quad
   + \[C_F^2\(\zeta_2-4\)+C_FC_A\(\frac{67}{36}-\frac{\zeta_2}2\) -\frac5{18}C_Fn_f\] \D_0(z_a) \D_0(z_b)
   \nonumber \\ & \quad
   + \[2\zeta_3C_F^2+C_FC_A\(\frac{11}{12}\zeta_2+\frac74\zeta_3-\frac{101}{54}\) +C_Fn_f \(\frac7{27}-\frac{\zeta_2}6\)\]
   \nonumber \\ & \qquad\times
                  \Big[\D_0(z_a) \delta(1-z_b) + \delta(1-z_a) \D_0(z_b)\Big]
   \nonumber \\ & \quad
   +\bigg\{ \D_2(z_a) F_2(z_b) + \D_1(z_a) F_1(z_b) + \D_0(z_a) F_0(z_b) + \delta(1-z_a)F_\delta(z_b)
   +(z_a\leftrightarrow z_b) \bigg\}
\end{align}
where the functions $F_i(z)$ are given by
\begin{subequations}
\begin{align}
  F_2(z) &= -\frac34C_F^2(1+z), \\
  F_1(z) &= C_F^2\(\frac12\frac{1+z^2}{1-z}\log\frac2{1+z} + \frac34(1+z)\log\frac{z}{(1-z)^2}-\frac{\log z}{1-z}\)
  \nonumber \\ & \quad
  +C_F\frac{11C_A-2n_f}{24}(1+z) + \frac{C_F}{4} \(\frac{1-z}2+\frac23 \frac{1-z^3}{z} +(1+z)\log z\), \\
  F_0(z) &= C_F^2\bigg[-\frac34(1+z)\log^2(1-z) + \(\frac{z^2}{1-z}\log\frac2{1+z}-\frac{3+z^2}2\frac{\log z}{1-z}\)\log(1-z)
    \nonumber \\ & \qquad\quad
    +\frac{1+15z}4+\(-1-\frac z2+\frac34z^2\)\frac{\log z}{1-z}+\frac{3+z^2}8\frac{\log^2z}{1-z}-\frac14\frac{1+z^2}{1-z}\log^2\frac{2z}{1+z}
    \nonumber \\ & \qquad\quad
    +\frac{1+z}2\(\Li_2\(\frac{1+z}2\)+\Li_2\(\frac{z-1}{2z}\)+\frac32\Li_2(-z)-\frac32\Li_2(z)+\frac{\zeta_2}4+\log2\log\frac{2z\sqrt z}{1+z}\) \bigg]
  \nonumber \\ & \quad\
  +C_FC_A\bigg[\frac{11}{24}(1+z)\log(1-z)+\frac5{18}-\frac{77}{36}z+\frac{\zeta_2}4(1+z) + \frac{17+5z^2}{24}\frac{\log z}{1-z}
    \nonumber \\ & \qquad\qquad\qquad
    + \frac{1+z^2}8\frac{\log^2 z}{1-z} -\frac{11}{24}\frac{1+z^2}{1-z}\log\frac2{1+z}  \bigg]
  \nonumber \\ & \quad
  +C_Fn_f\bigg[-\frac1{12}(1+z)\log(1-z)+\frac1{18}+\frac29z+\frac1{12}\frac{1+z^2}{1-z}\(\log\frac2{1+z}-\log z\)\bigg]
  \nonumber \\ & \quad
  +C_F\bigg[\frac{25+7z+22z^2}{36z}(1-z)
    -\frac{1+z}z\(\frac23+\frac5{24}z+\frac{z^2}6\)\log\frac2{1+z} + \(\frac18+\frac58z+\frac{z^2}3\)\log z
    \nonumber \\ & \qquad\qquad
    +\frac{1+z}4\(\Li_2(1-z)+\Li_2(-z)+\frac{\zeta_2}2+\log2\log z-\frac12\log^2z\)  \bigg], \\
  F_\delta(z) &= C_F^2\(\frac74\zeta_2-2\)\(1-z-(1+z)\log(1-z)+\frac{1+z^2}{1-z}\log\frac2{1+z}\)
   \nonumber \\ & \quad
                  +\frac1{16}\text{reg}\[\tilde I^{(2)}_{qqV}(z) + \tilde I^{(2)}_{qqS}(z)\].
\end{align}
\end{subequations}
In the last function we have kept the dependence on the functions $I^{(2)}_{qqV}(z)$ and $I^{(2)}_{qqS}(z)$
given in equation (S53) of Ref.~\cite{Lustermans:2019cau}. Specifically,
these functions contain distributional terms in $\D_k(z)$ and $\delta(1-z)$,
which are already taken into account explicitly in Eq.~\eqref{eq:C2tildeLMT}
as they contribute to the double-distributional part of the result,
so here only the remaining regular part, denoted by ${\rm reg}\[...\]$ in the formula, has to be considered.
The factor $1/16$ finally fixes the different normalization due to the different
expansion parameters (we use $\as/\pi$ while LMT use $\as/(4\pi)$).
It is useful to write explicitly the large $z$ expansion of the $F_i(z)$ functions,
\begin{subequations}
\begin{align}\label{eq:C2tildeLMTexp}
  F_2(z) &= -\frac32C_F^2, \\
  F_1(z) &= -3C_F^2\log(1-z) + \frac32C_F^2 +C_F\frac{11C_A-2n_f}{12} , \\
  F_0(z) &= -\frac32C_F^2\log^2(1-z)+\(\frac52C_F^2+C_F\frac{11C_A-2n_f}{12}\)\log(1-z)
   \nonumber \\ & \quad
                  +C_F^2\(\frac{19}4-\zeta_2\) +C_FC_A\(-\frac{233}{72}+\frac{\zeta_2}2\)+C_Fn_f\frac{19}{36}, \\
  F_\delta(z) &= -\frac12C_F^2\log^3(1-z)+ \(\frac32C_F^2+C_F\frac{11C_A-2n_f}{24}\)\log^2(1-z)
   \nonumber \\ & \quad
                  + \(C_F^2\(\frac{33}8-\zeta_2\) +C_FC_A\(-\frac{47}{18}+\frac{\zeta_2}2\)+C_Fn_f\frac{19}{36}\)\log(1-z)
   \nonumber \\ & \quad
                  + C_F^2\(-2-\frac{\zeta_2}2-2\zeta_3\) +C_FC_A\(\frac{125}{54}-\frac76\zeta_2-\frac74\zeta_3\)+C_Fn_f\(-\frac{67}{108}+\frac{\zeta_2}6\).
\end{align}
\end{subequations}
From this expansion it is easy to verify that the double-Mellin transform of the LMT result coincides at next-to-leading power with
the AMRST result Eq.~\eqref{eq:C2tildeBDDR}.

\subsection{Analytical comparison}
\label{sec:appcomparison}

We consider here an analytical comparison of the approaches studied in this work.
We find it convenient to perform this comparison in double Mellin space,
which allows us to obtain more compact expressions.
The expressions for BDDR and AMRST are already given in this space, Eqs.~\eqref{eq:C1tildeBDDR} and \eqref{eq:C2tildeBDDR},
and LMT can be easily computed by a direct double Mellin transform of Eqs.~\eqref{eq:C1tildeLMT} and \eqref{eq:C2tildeLMT}.
As far as BNX and BFR are concerned, the computation of the double Mellin tranform is
less obvious as they are expressed in terms of the $z,u$ variables.
Using the definitions of section~\ref{sec:thr}, it is easy to prove that
\begin{align}
  \tilde C(N_a,N_b)
  &\equiv \int_0^1dz_a\, z_a^{N_a-1} \int_0^1dz_b\, z_b^{N_b-1}\,\tilde C(z_a,z_b)\nonumber\\
  &= \int_0^1dz\,z^{N-1} \int_0^1du\, \(\frac{1-(1-z)u}{z+(1-z)u}\)^{\Delta N}
    \frac{z(1+z)^2 C(z,u)}{\(1-(1-z)u\) \(z+(1-z)u\)(z_a+z_b)^2}
\end{align}
with $N=\frac{N_a+N_b}2$ and $\Delta N=\frac{N_a-N_b}2$, and $z_{a,b}(z,u)$ given in Eq.~\eqref{eq:zazbdef}.
Plugging in the expressions of $C(z,u)$ Eqs.~\eqref{eq:BNX} and \eqref{eq:BFR} for BNX and BFR respectively,
we find
\begin{align}
  \tilde C^{\rm BNX}(N_a,N_b) &= \frac12\Big[C_{\rm thr}(N_a) + C_{\rm thr}(N_b)\Big], \label{eq:BNXmell}\\
  \tilde C^{\rm BFR}(N_a,N_b) &= C_{\rm thr}\(\frac{N_a+N_b}2\). \label{eq:BFRmell}
\end{align}
We also recall that $\tilde C(N_a,N_b)$ computed in $N_a=N_b$ corresponds to the Mellin transform
of the rapidity-integrated coefficient function $C(z)$.
Indeed, inserting $1=\int_0^1dz\,\delta(z-z_az_b)$ in the definition of the double Mellin transform, we get
\begin{align}
  \tilde C(N,N)
  &= \int_0^1 dz\, z^{N-1} \int_0^1dz_a \int_0^1dz_b\,\delta(z-z_az_b)\tilde C(z_a,z_b) \nonumber\\
  &= \int_0^1 dz\, z^{N-1} C(z) \nonumber\\
  &= C(N).
\end{align}
Eqs.~\eqref{eq:BNXmell} and \eqref{eq:BFRmell} clearly satisfy this condition, modulo the fact
that the rapidity-integrated coefficient is approximated at threshold.

In the following, we will ignore BFR
(which is equivalent to BNX up to next-to-next-to-leading power in
$1-z$ as we have already commented in section~\ref{sec:proof})
and consider the BNX expression only, Eq.~\eqref{eq:BNXmell},
computed with the $\psi$-soft$_1$ choice of threshold logarithms,
as described in section~\ref{sec:approx}.
We first perform the comparison at the rapidity-integrated level, where we find
at NLO and NNLO
\begin{align}
  C_1(N) &= \tilde C_1^{\rm BNX,\, \psi\text{-soft}_1}(N,N) + \Ord\(\frac1{N^2}\) \\
  C_2(N) &= \tilde C_2^{\rm BNX,\, \psi\text{-soft}_1}(N,N)\nonumber\\
         &\quad + \frac1N  \[ \frac72C_F^2L^2 + \(-\frac74C_F^2+\frac{29}{12}C_FC_A-\frac23C_Fn_f\)L -\frac{\zeta_2}2C_F^2- \frac{35}{36}C_FC_A-\frac49C_Fn_f\]\nonumber\\
         &\quad + \Ord\(\frac1{N^2}\), \label{eq:C2BNXdiffint}
\end{align}
with $L=\log N+\gamma$.
We thus see what we expected from $\psi$-soft$_1$, namely that on top of the leading power contributions
also the LL term at next-to-leading power is predicted correctly.
At NLO, the next-to-leading power NLL term, of the form $1/N$, is also correctly predicted by $\psi$-soft$_1$,
as well as part of next-to-leading power NLL terms at NNLO, of the form $L^2/N$
(in particular, the term proportional to $\beta_0$ in the fourth line of Eq.~\eqref{eq:C2tildeBDDR} is completely reproduced).

Moving to the rapidity-differential coefficient, i.e.\ keeping $N_a\neq N_b$, we find
\begin{align}
  \tilde C_1(N_a,N_b) &= \tilde C_1^{\rm BNX,\, \psi\text{-soft}_1}(N_a,N_b) -\frac{C_F}2\dL^2 -\frac{C_F}2\frac{\dL}{\dN} + \Ord\(\frac1{\bar N^2}\) \label{eq:C1BNXdiff}\\
  \tilde C_2(N_a,N_b) &= \tilde C_2^{\rm BNX,\, \psi\text{-soft}_1}(N_a,N_b)\nonumber\\
         &\quad -C_F^2\(\frac18\dL^4+ \frac34\Lb^2\dL^2\) - C_F\frac{11C_A-2n_f}{24}\Lb\dL^2\nonumber\\
         &\quad + \(C_F^2(2-2\zeta_2)+C_FC_A\(-\frac{67}{72}+\frac{\zeta_2}4\)+C_Fn_f\frac5{36}\)\dL^2\nonumber\\
                      &\quad + \frac1{\bar N}\[\frac7{16}C_F^2\Lb^2+\(-\frac7{16}C_F^2+\frac{29}{48}C_FC_A-\frac{C_Fn_f}6\)\Lb
                        -\frac{\zeta_2}4C_F^2- \frac{35}{72}C_FC_A-\frac29C_Fn_f\]\nonumber\\
         &\quad -\frac{C_F^2}4\(\frac{\dL^3}{\dN}+3\frac{\Lb^2\dL}{\dN}+3\frac{\Lb\dL^2}{\bar N}\) \nonumber\\
                      &\quad + \(\frac38C_F^2-C_F\frac{11C_A-2n_f}{24}\)\frac{\Lb\dL}{\dN}
                        - \(\frac{C_F^2}{16}+C_F\frac{11C_A-2n_f}{48}\)\frac{\dL^2}{\bar N} \nonumber\\
                      &\quad + \(C_F^2\(\frac{37}{16}-2\zeta_2\)+C_FC_A\(-\frac{179}{144}+\frac{\zeta_2}4\)+C_Fn_F\frac5{36}\) \frac{\dL}{\dN}\nonumber\\
                      &\quad + \Ord\(\frac1{\bar N^2}\),
                        \label{eq:C2BNXdiff}
\end{align}
with
\begin{align}\label{eq:dLdN}
  \frac1{\dN} &\equiv \frac1{N_a} - \frac1{N_b} &
  \dL &\equiv \Lb_a-\Lb_b=\log\frac{N_a}{N_b},
\end{align}
and having used the definitions Eq.~\eqref{eq:LbarNbar}.
We immediately notice that, in addition to the contributions already present at the integrated level,
i.e.\ the fourth line of Eq.~\eqref{eq:C2BNXdiff},
the BNX result differs from the exact (and equivalently from AMRST) by many other terms,
appearing also at leading power, namely without a $1/N_{a,b}$ suppression.
These terms are all proportional to at least one power of $\dL$,
and therefore vanish in the $N_a=N_b$ limit that reproduces the rapidity-integrated result.

The presence of these contributions already at leading power may seem worrisome.
However, we have already encountered contributions that are apparently leading power at parton level
but that contribute at next-to-leading power to the cross section: we have discussed them in section~\ref{sec:leadingpower}.
There, we had shown that contributions of the form Eq.~\eqref{eq:Gdef} vanishing after integration over $u$
contribute at next-to-leading power to the cross section.
Even more, if the contributions are symmetric under the exchange $u\to1-u$ (namely are symmetric in partonic rapidity),
then their contribution is even more suppressed, at next-to-next-to-leading power level.
Since integrating over $u$ corresponds to integrating over partonic rapidity,
all leading-power contributions in $\tilde C(N_a,N_b)$ that vanish when $N_a=N_b$
belong to the category discussed in section~\ref{sec:leadingpower}.
Moreover, in Eqs.~\eqref{eq:C1BNXdiff} and \eqref{eq:C2BNXdiff}, all terms proportional to $\dL$
are symmetric under the exchange $N_a\leftrightarrow N_b$, corresponding to a sign flip of $\dL$ and $\dN$,
which means that they are the Mellin transform of terms symmetric under the exchange $u\to1-u$.
Therefore, the difference between BNX and AMRST at the cross section level
is given at next-to-leading power just by the terms appearing also in the rapidity-integrated coefficient Eq.~\eqref{eq:C2BNXdiffint},
while the additional terms proportional to $\dL$ showing up in Eqs.~\eqref{eq:C1BNXdiff} and \eqref{eq:C2BNXdiff}
only contribute at next-to-next-to-leading power to the cross section.
This explains the good agreement between the two formulations.

As far as LMT is concerned, we have already commented at the end of section~\ref{sec:LMT}
that up to next-to-leading power it coincides with AMRST, the difference starting at next-to-next-to-leading power.
We thus do not need to give explicit $N$-space expressions here.

\phantomsection
\addcontentsline{toc}{section}{References}

\bibliographystyle{jhep}
\bibliography{references}

\end{document}